\documentclass{LMCS}

\def\dOi{10(3:7)2014}
\lmcsheading%
{\dOi}
{1--21}
{}
{}
{Jan.~13, 2013}
{Aug.~19, 2014}
{}

\ACMCCS{[{\bf Mathematics of computing}]: Continuous mathematics}

\usepackage{amssymb}
\usepackage{amsmath}
\usepackage{latexsym}

\newcommand{\hide}[1]{}

\def \dom{{\rm dom}}

\def \id{{\rm id}}

\def \In{{\subseteq}}
\def \s{{\Sigma^*}}
\def \om{{\Sigma^\omega }}
\def \pf{:\hspace{0.6ex}\subseteq \hspace{-0.4ex}}

\def\IN{{\mathbb{N}}}
\def\IQ{{\mathbb{Q}}}
\def\IR{{\mathbb{R}}}

\def\EE{{\mathcal E}}
\def\AA{{\mathcal A}}
\def\RR{{\mathcal R}}
\def\XX{{\mathcal X}}
\def\YY{{\mathcal Y}}

\def\syd{{\,\Delta\,}}
\newcommand{\an}{\ \ \mbox{and}\ \ }
\newcommand{\qq}{\qed}
\newcommand{\bb}{ \hspace{-0.76ex}- \hspace{-0.80ex}}

\newcommand{\mto}{\rightrightarrows}

\begin{document}

\title[Computability on measurable sets]{Representations of measurable sets in computable measure theory }

\author[K.~Weihrauch]{Klaus Weihrauch\rsuper a} 
\address{{\lsuper a}Department of Mathematics and Computer Science, University of Hagen, Germany}    
\email{Klaus.Weihrauch@FernUni-Hagen.de}  

\author[N.~R.~Tavana]{Nazanin Roshandel Tavana\rsuper b}
\address{{\lsuper b}Department of Mathematics and Computer Science, Amirkabir
University of Technology, and
School of Mathematics, Institute for Research in Fundamental
Sciences (IPM), Tehran, Iran}
\email{nazanin.r.tavana@ipm.ir}
\thanks{{\lsuper b}The second author was partially supported by a grant from IPM, grant number 92030118}



\begin{abstract}
This article is a fundamental study in computable measure theory. We
use the framework of TTE, the representation approach, where
computability on an abstract set $X$ is defined by representing its
elements with concrete ``names'', possibly countably infinite, over
some alphabet $\Sigma$.  As a basic computability structure we consider
a computable measure on a computable $\sigma$-algebra. We introduce
and compare w.r.t. reducibility several natural representations of
measurable sets. They are admissible and generally form four different
equivalence classes. We then compare our representations with those
introduced by Y. Wu and D. Ding in 2005 and 2006 and claim that one of
our representations is the most useful one for studying computability
on measurable functions.
\end{abstract}

\maketitle

\section{Introduction}\label{seca}
Measure theory is a fundament of modern analysis. In particular, computable
measure theory is a fundament of computable analysis. In recent years a number of articles have been published on computable measure theory, for example \cite{Eda95,Mue99,Wei99a,Gac05,WW06,Sch07b,Bos08b,Eda09,HR09,GHR09b,WWD09,GHR10,AFR11,GHR11,HRW11c,BGHRS11,FR12,Wu12,BDHMS12,MTY13,Hoy13}. Most of these articles start with a definition of computability concepts in measure
theory and then prove, or disprove, a computable version of some classical theorem.

Wu and Ding  \cite{WD05,WD06} have defined  and compared various  definitions of computability on measurable sets. In this article we extend these fundamental studies.  We use the representation approach to computable analysis (TTE) \cite{Wei00,BHW08}. In this approach computability is defined directly on the set $\om$ of the infinite sequences of symbols, e.g. by Turing machines. Computability is transferred to other sets $X$ by means of representations
$\delta:\om\to X$ where the elements of $\om$ are considered as names and computations are performed on names. Obviously, computability on the ``abstract"  set $X$ depends crucially on the choice of the representation $\delta$. Only those representations are of interest which can relate the important structure properties of $X$ with corresponding ones of $\om$.

We start from a computable measure on a computable $\sigma$-algebra which has proved to be a very useful fundamental concept of computability in measure theory \cite{WD05,WD06,WW06}.
In addition to the representations studied in these articles we introduce several new representations of the measurable sets and compare all of them w.r.t. reducibility.

In Section~\ref{secf} we outline very shortly some concepts from the representation approach.
In Section~\ref{secb} we summarize elementary definitions and facts from measure theory which we will need for introducing the new computability concepts.

In Section~\ref{secc} we define computable $\sigma$-algebras $(\Omega,\AA,\RR,\alpha)$
where $\RR$ is a countable ring which generates the $\sigma$-algebra $\AA$ in $\Omega$
such that $\Omega=\bigcup \RR$ and $\alpha\pf \s\to \RR$  is a notation of the ring such that set union and difference become computable. A measure $\mu$ is computable if $\mu(R)$ is finite for every ring element $R$ and $R\mapsto\mu(R)$ is computable.
Then we introduce and study representations  $\zeta_+$, $\zeta_-$ and $\zeta$ of the measurable sets  which exactly allow to compute $\mu(R\cap A)$ for for $R\in\RR$ and $A\in\AA$ from below, from above or from below and above, respectively. We study reducibility and characterize the degree of non-computability for the negative results.

In Section~\ref{sece} for the sets of finite measure we define a computable metric space and compare its Cauchy representation with the representations defined before.

In Section~\ref{secd} we partition the set $\Omega$ computably by a (majorizing) sequence $(F_i)_{i\in\IN}$ of ring elements. For each number $i$, the measure restricted to $F_i$ is finite and induces a computable metric space, the metric of which can be normalized to a metric $d_i'$ bounded by $1$. The weighted sum $\overline d=\sum_i 2^{-i}\cdot d'_i$ is a computable metric on the whole $\sigma$-algebra the Cauchy representation of which allows to compute the measures of measurable sets from below and above and hence  is equivalent to the representation $\zeta$ from Section~\ref{secc}.

In Section~\ref{secg} we show that all the representations are admissible \cite{Wei00}. We compare our representations with those from \cite{WD05,WD06}.
It turns out that $\zeta_+$ for which there is no equivalent one in \cite{WD05,WD06} is most interesting.

\section{Computability by means of representations}\label{secf}

For studying computability we use the TTE, representation approach to computable analysis \cite{Wei00,BHW08}. Let $\Sigma$ be a fixed finite alphabet such that $0,1\in \Sigma$. $\s $ denotes the set of finite words over $\Sigma$ and $\om$ denotes the set of infinite sequences $p:\IN\to\Sigma$. A partial function $f\pf Y_1\times\ldots Y_k\to Y_0$ (where $Y_i=\s$ or $Y_i=\om$) is computable, iff it can be computed by a Type-2 Turing machine. For encoding pairs and longer tuples of elements from $\s$ and $\om$ we use tupling functions all of which are denoted by $\langle\ \rangle$  \cite[Definition~2.1.7]{Wei00}. For  the wrapping function
$\iota:\s\to\s$, $\iota(a_1a_2\ldots a_k):=110a_10a_2\ldots0a_k011$, two wrapped words cannot overlap properly. For $ w_i\in \s$ and $p_i\in \om$ let $\langle w_1,\ldots,w_n\rangle:=\iota(w_1)\iota(w_2)\ldots\iota(w_n)$,
$\langle w_0,p_0\rangle :=\langle p_0,w_0\rangle :=  \iota(w_0)p_0\in\om$,
$\langle p_0,p_1\rangle :=(p_0(0)p_1(0)p_0(1)p_1(1)\ldots)$,
$\langle p_0,p_1,\ldots\rangle\langle i,j\rangle:=p_i(j)$ (where $\pi:\IN^2\to\IN$, $\langle i,j\rangle=\pi(i,j)$ is a standard computable bijection),
 etc. The tupling functions  and the projections of their inverses are computable.
We will use definitions of the form ''$p$ is a list of all pairs $(u,v)\in\s\times\s$ such that $Q(u,v)$'' meaning: $\iota(\langle u,v\rangle)$ is a subword of $p$ iff $Q(u,v)$.

We use canonical representations  $\nu_\IN\pf \s\to \IN$, $\nu_\IQ\pf \s\to\IQ$,
of the natural numbers and the rational numbers, respectively. For the real numbers let
$\rho_<(p)=x$ iff $p$ is a list of all $u$ such that $\nu_\IQ(u)<x$,
$\rho_>(p)=x$ iff $p$ is a list of all $u$ such that $\nu_\IQ(u)>x$ and $\rho\langle p,q\rangle=x$ iff $\rho_<(p)=x$ and $\rho_>(q)=x$. The representations $\overline \rho_<$,
$\overline \rho_>$, $\overline \rho$ of the set $\overline \IR:= \IR\cup\{-\infty,\infty\}$ are defined accordingly \cite[Section~4.1]{Wei00}.

A representation of a set $X$ is a partial surjective function $\delta\pf Y\to X$ where $Y=\s$ or $Y=\om$. For representations $\delta_i\pf Y_i\to X_i$, ($i=1,2$), a function $h\pf Y_1\to Y_2$ (operating on names) realizes the (abstract) function $f\pf X_1\to X_2$, iff $f\circ \delta_1(p)=\delta_2\circ h(p)$ for all
$p\in\dom(f\circ \delta_1)$. A function $f$ is called $(\delta_1,\delta_2)$-continuous (-computable), iff it is realized by a continuous (computable) function. A representation $\delta_1$ is reducible to (translatable to) $\delta_2$, $\delta_1\leq\delta_2$,  iff the identity function ${\rm id}:x\mapsto x$ is $(\delta_1,\delta_2)$-computable, that is, there is a computable function $h$ such that $\delta_1(p)=\delta_2\circ h(p)$ for all $p\in\dom(\delta_1)$. Correspondingly, $\delta_1$ is topologically  reducible to $\delta_2$, $\delta_1\leq_t\delta_2$, iff there is a continuous function $h$ such that $\delta_1(p)=\delta_2\circ h(p)$ for all $p\in\dom(\delta_1)$.
The two representations are equivalent, $\delta_1\equiv\delta_2$, iff $\delta_1\leq\delta_2$ and  $\delta_2\leq\delta_1$. Accordingly, they are topologically equivalent,  $\delta_1\equiv_t\delta_2$, iff $\delta_1\leq_t\delta_2$ and  $\delta_t\leq_t\delta_1$.
Equivalent representations induce the same computability on the represented sets.
For more details see \cite{Wei00,BHW08}.

\section{Concepts from classical measure theory}\label{secb}

In this Section we summarize elementary definitions and facts from measure theory which we will need for introducing the new computability concepts.

Let $\Omega$ be a set. \\
-- A ring  (in $\Omega)$ is a set $\RR\In 2^\Omega$  such that $\emptyset\in\RR$, and $A\cup B\in \RR$ and $A\setminus B\in\RR$ if $A,B\in\RR$. Since $A\cap B = A\setminus (A\setminus B )$, every ring is closed under intersection. The ring is called an algebra, if $\Omega\in\RR$.\\
-- A $\sigma$-algebra (in $\Omega$) is a set $\AA\In 2^\Omega$ such that $\Omega\in \AA$, $A^c=\Omega\setminus A\in \AA$ if $A\in\AA$, and $\bigcup_{i=0}^\infty A_i\in\AA$ if $A_0,A_1,\ldots\in\AA$. The elements of $\AA$ are called the measurable sets. Every $\sigma$-algebra is a ring.\\
-- For a set ${\mathcal T} \In {2^\Omega}$, $\RR({\mathcal T})$ denotes the smallest ring containing $\mathcal T$ and $\AA({\mathcal T})$ denotes the smallest $\sigma$-algebra containing $\mathcal T$.\\
-- A measure on a ring $\RR$ is a function $\mu:\RR\to \IR^\infty$ ($=\IR\cup \{\infty\}$) such that
$\mu(\emptyset) =0$, $\mu(A)\geq 0$ for all $A\in\RR$, and $\mu(\bigcup_iA_i)=\sum_i\mu(A_i)$ for pairwise disjoint sets $A_0,A_1,\ldots\in\RR$ such that $\bigcup_iA_i\in \RR$. (Often $\mu$ is called a pre-measure if $\RR$ is a ring and a measure only if $\RR$ is a $\sigma$-algebra.)\\
--  A measure $\mu$ on a ring $\RR$ is $\sigma$-finite, if there is a sequence $E_0,E_1,\ldots\in\RR$ of sets such that $ (\forall i) \mu(E_i)<\infty \ \ \mbox{and}\ \  \bigcup_i E_i=\Omega$. The sets can be assumed to be pairwise disjoint: for $F_j:= E_j\setminus \bigcup _{i<j} E_i$ the $F_j$ are ring elements such that
\begin{eqnarray}\label{f4}
 (\forall i\neq j)\, F_i\cap F_j=\emptyset, \ \ (\forall i) \mu(F_i)<\infty \ \ \mbox{and}\ \  \bigcup_i F_i=\Omega \,.
\end{eqnarray}

For two sets $A,B$ let $A\syd
B:= (A\setminus B) \cup (B\setminus A)$ be their symmetric difference. Some useful rules for the symmetric difference are listed in the appendix Section~\ref{sech}.

Two sets $A$ and $B$ with $\mu(A\syd B)=0$ are essentially identical in measure theory.

\begin{defi}\label{d6} Let $\mu$ be a measure on a ring \,$\RR$. Define an equivalence relation on \,$\RR$ by
$A\sim B \iff \mu(A\syd B)=0$. Let  $[A] := \{B\in\AA\mid A\sim B \}$ be the equivalence class containing $A$.
For ${\mathcal E}\In \RR$, let
$[{\mathcal E}]:=\{[A]\mid A\in {\mathcal E}\}$.
\end{defi}

Notice that the following are equivalent: $\mu(A\syd B)=0$, \  $A\sim B$ , $A\in [B]$, $B\in [A]$, and $[A]=[B]$.

\begin{lem}\label{l9} For $i\in\IN$ let $A_i,B_i\in\RR$
such that $A_i\sim B_i$.\\ Then $\mu(A_0)=\mu(B_0)$,
$A_0 {\;\rm op\;} A_1 \sim B_0{\;\rm op\;} B_1$ for \:${\rm op}\in\{\cup,\cap,\setminus\}$,
 and \,$\bigcup_i A_i\sim \bigcup_i B_i$ if
\ $\bigcup_i A_i\in\RR$ and \, $\bigcup_i B_i\in\RR$.

Therefore the following operations are well-defined on equivalence classes:\\
$\mu([A_0]):= \mu(A_0)$, $[A_0] {\;\rm op\;} [A_1]:=[A_0 {\;\rm op\;} A_1]$ for \:${\rm op}\in\{\cup,\cap,\setminus\}$,
and $\bigcup_i[A_i]:=[\bigcup_iA_i]$ if \,$\bigcup_i A_i\in\RR$.
\end{lem}

\proof
Straightforward, using in particular (\ref{f10}) and (\ref{f11}).
\qq

Our computability concepts in measure theory are based on the following theorem.

\begin{thm}[Carath\'eory extension theorem \cite{Bau74,Fol99}]\label{t1}
Every $\sigma$-finite measure on a ring $\RR$ has a unique extension to a measure on the $\sigma$-algebra $\AA(\RR)$.
\end{thm}

Therefore, for specifying a measure $\mu$ on the $\sigma$-algebra $\AA(\RR)$, it suffices to define $\mu(E)$ for every $E\in \RR$.

Let $\mu$  be a $\sigma$-finite measure on a ring $\RR$ and let $(F_i)_{i\in\IN}$ be a sequence of ring elements which satisfy (\ref{f4}).
For any set $R\in\RR$, we have $\mu(R)=\sum_{i\in\IN}\mu(R\cap F_i)$. This implies that the measure $\mu$ on the ring $\RR$ is completely determined by its restriction to the subring $\RR^f$ which consists of all ring elements with finite measure.

In our computable measure theory we will consider only  $\sigma$-algebras $\AA(\RR)$ spanned by a finite or countable ring $\RR$ (which is non-empty since $\emptyset\in\RR$) and measures $\mu$ such that $\mu(R)<\infty$ for all $R\in\RR$ and $\mu$ is $\sigma$-finite on~$\RR$.

\begin{lem}\label{l3}
Let $\mu$ be a measure on a countable ring $\RR\In 2^\Omega$ such that $\mu(R)<\infty$ for all $R\in\RR$.
\begin{enumerate}
\item \label{l3a} If \ $\bigcup \RR =\Omega$ then the measure $\mu$ is \, $\sigma$-finite.
\item \label{l3b}  The measure $\mu$ is $\sigma$-finite in $\Omega':=\bigcup \RR$.
\end{enumerate}
\end{lem}

\proof

(\ref{l3a}) Since $\RR\neq \emptyset$ it has an enumeration (not necessarily injective) $(E_i)_{i\in\IN}$. Let $F_0:=E_0$ and $F_{n+1}:=E_{n+1}\setminus (E_0\cup\ldots\cup E_n)$. Then the sets $F_i$ are pairwise disjoint elements of $\RR$ such that $\mu(F_i)<\infty$ and $\bigcup_iF_i=\bigcup_iE_i=\bigcup\RR=\Omega$.

(\ref{l3b}) Since $\RR\In 2^{\Omega'}$, $\mu$ is $\sigma$-finite in $\Omega'$ by~(\ref{l3a}).
\qq

Therefore, if $\bigcup\RR\neq\Omega$, we obtain a $\sigma$-finite measure by ignoring $\Omega\setminus \bigcup\RR$.
We will use the next two theorems for defining representations of the measurable sets.
For a measure $\mu$ on a $\sigma$-algebra $\AA$ and a subset $\EE\In\AA$ let $\EE^f:=\{A\in  \EE \mid \mu(A)<\infty\}$ be the set of elements of $\EE$ of finite measure.

Special cases of the following theorem are proved in most introductory texts. A complete proof is added in the appendix Section~\ref{seci}.

\begin{thm}\label{t2}
Let $\mu$ be a measure on a $\sigma$-algebra $\AA$. On $ \AA^f$ the Fr\'echet metric is defined by $d(A,B):= \mu(A\syd B)$.
\begin{enumerate}
\item \label{t2a} $({\AA}^f,d)$, is a complete pseudometric space.
\item \label{t2b}
Let $(A_i)_{i\in\IN}$ be a sequence in $ \AA^f$ such that $d(A_i,A_j)\leq 2^{-i}$ for $j>i$. \\
For $m\leq k$ let $B_{mk}:=\bigcup_{i=m}^k A_i$, let
$B_m:=\bigcup _{i\geq m}A_i$ and \\
$B:=\bigcap_m B_m= \bigcap_m\bigcup_{i\geq m}A_i$. Then
\begin{eqnarray}
\label{f21}
&&\hspace{-9ex}B_{mk}\In B_{m,k+1},  \ d(B_{mk}, B_{m,k+1})\leq 2^{-k}
 \mbox{ and }\  d(B_{mk}, B_m)\leq2\cdot 2^{-k},\\
\label{f22}
&&\hspace{-9ex}B_m \supseteq B_{m+1}\in\AA^f, \
    \ d(B_{m}, B_{m+1})\leq 2^{-m}\mbox{ and }\
     d(B_m, \ B)\leq 2\cdot 2^{-m}, \\
\label{f23}
&&\hspace{-9ex}B\in\AA^f\mbox{ and } d(A_m,B)\leq 4\cdot 2^{-m}\,.
\end{eqnarray}

\item \label{t2d}
Let $(A_i)_{\in\IN}$ be a sequence in $ \AA^f$ such that $d(A_i,A_j)\leq 2^{-i}$ for $j>i$.\\
For $m\leq k$ let $D_{mk}:=\bigcap_{i=m}^k A_i$, \
let $D_m:=\bigcap _{i\geq m} A_i$ and \\
$D:=\bigcup_m D_m =\bigcup_m\bigcap_{i\geq m} A_i$. Then
 \begin{eqnarray}
 \label{f25}
 &&\hspace{-9ex}D_{mk}\supseteq D_{m,k+1}, \ d(D_{mk}, D_{m,k+1})\leq 2^{-k}  \mbox{ and }\  d(D_{mk}, D_m)\leq 2\cdot 2^{-k},\\
 \label{f26}
 &&\hspace{-9ex}D_m \In D_{m+1}\in\AA^f,
      \ d(D_{m}, D_{m+1})\leq 2^{-m}\mbox{ and }\  d(D_m, \ D)\leq 2\cdot 2^{-m}, \\
 \label{f27}
 &&\hspace{-9ex}D\in\AA^f\mbox{ and } d(A_m,D)\leq 4\cdot 2^{-m}\,.
 \end{eqnarray}


\item \label{t2c} If \,$\RR$ is a ring such that $\AA:=\AA(\RR)$ and the measure $\mu$ is
$\sigma$-finite on $\RR$, then $\RR^f$ is a dense subset of $\AA^f$.

\end{enumerate}
\end{thm}

\noindent If $d(A_i,A_j)\leq 2^{-i}$ for $j>i$ then by Theorem~\ref{t2} the sequence $(A_i)_{i\in\IN}$ converges to $B=\bigcap_m\bigcup_{i\geq m}A_i\in \AA^f$ and to  $D=\bigcup_m\bigcap_{i\geq m}A_i\in \AA^f$. Notice that $\bigcup_m\bigcap_{i\geq m}A_i  \;\In\; \bigcap_m\bigcup_{i\geq m}A_i$
since ($x\in A_i$ for almost all $i$) implies ($x\in A_i$ infinitely often)
and that $\mu(\bigcap_m\bigcup_{i\geq m}A_i \setminus \bigcup_m\bigcap_{i\geq m}A_i)=0$.

\medskip

A set $A\in\AA$ is determined uniquely up to a set of measure $0$ by the values $\mu(A\cap E)$ for ring elements $E$ of finite measure. We will use this fact for defining various representations of the set $[\AA]$.

\begin{lem}\label{l6}
Let $\RR$ be a ring  and let $\mu$ be a measure on $\AA:=\AA(\RR)$ which is
$\sigma$-finite on $\RR$. Then for $A,B\in\AA$,
\[ \mu(A\syd B)=0\iff (\forall E\in\RR^f)\, \mu(A\cap E)=\mu(B\cap E)\,.\]
\end{lem}

\proof 
$\Longrightarrow :$ Suppose $\mu(A\syd B)=0$ and $E\in \RR$. Then \\
$\mu((A\cap E)\syd (B\cap E))=\mu((A\syd B)\cap E)=0$, hence $\mu(A\cap E)=\mu(B\cap E)$ by (\ref{f1}).
 \medskip

\noindent $\Longleftarrow :$
Suppose  $\mu(A\syd B)>0$. We may assume, without loss of generality, $\mu(A\setminus B)>0$.
 We want to find some $E\in \RR^f$ such that $\mu(A\cap E)\neq \mu(B\cap E)$.
Since $\Omega=\bigcup_{R\in\RR}R$,

$\mu(A\setminus B)=\mu(\bigcup _{R\in\RR}R\cap(A\setminus B))
=\mu(\bigcup _{R\in\RR}(R\cap(A\setminus B)))
\leq\sum_{R\in\RR}\mu((R\cap A)\setminus (R\cap B)) $.\\
Therefore, $c:=\mu((R\cap A)\setminus R\cap B))>0$ for some $R\in\RR$.

Let $C:= R\cap  A$ and $D:=  R\cap B$. Then $\mu(C\setminus D)=c$.
By Theorem~\ref{t2}(\ref{t2c}) there is some $G\in\RR^f$ such that
$\mu(G\syd(C\setminus D))< c/3 $.  By (\ref{f9}),
$$D\cap G\In D\cap((C\setminus D) \cup (G\syd(C\setminus D)))
\In G\syd(C\setminus D)\,,$$
 hence $\mu(D\cap G)<c/3$.
Again by (\ref{f9}),  $C\setminus D\In G\cup (G\syd(C\setminus D))$, hence
$$C\setminus D=C\cap(C\setminus D)\In (C\cap G)\cup (G\syd(C\setminus D))\,,$$
therefore, $c=\mu(C\setminus D)\leq \mu(C\cap G) + c/3$.
We obtain $\mu(B\cap R\cap G) = \mu(D\cap G)<c/3$ and
$\mu(A\cap R\cap G)=\mu(C\cap G)\geq 2c/3$, hence for $E:= R\cap G$, $\mu(A\cap E)\neq \mu(B\cap E)$.
\qq

\section{The basic representations}\label{secc}

In computable analysis computability on an uncountable structure is usually introduced by selecting a countable substructure which ``generates'' it
and defining the meaning of ``computable'' on this substructure (example: computability on the field $\IQ$, completion to $\IR$). The results from the last section suggest that a countable ring with a $\sigma$-finite measure should be a good substructure. Then ring operations should become computable as well as the measure restricted to the ring.

\begin{defi}[Computable $\sigma$-algebra, computable measure]\label{d1}$ $\hfill
\begin{enumerate}
\item A {\em computable $\sigma$-algebra} is a tuple
$(\Omega,\AA,\RR,\alpha)$ such that $\RR$  is a countable ring in $\Omega$, $\Omega=\bigcup\RR$, $\AA=\AA(\RR)$, and
$\alpha\pf \s\to \RR$ is a notation of $\RR$ such that
 $\dom(\alpha)$ is recursive
and the functions $(A,B)\mapsto A\cup B$ and  $(A,B)\mapsto A\setminus B$ are computable (w.r.t. $\alpha$).
\item A measure $\mu$ on a computable $\sigma$\bb algebra $(\Omega,\AA,\RR,\alpha)$ is computable, if it is finite on $\RR$ and \  $R\mapsto\mu(R)$, the restriction of $\mu$ to $\RR$, is $(\alpha,\rho)$\bb computable.
\end{enumerate}
\end{defi}\smallskip

\noindent For a computable $\sigma$-algebra the intersection operation on the ring is also computable because $A\cap B = A\setminus (A\setminus B )$. Sometimes it is more convenient to use a numbering $E:\IN\to \RR $ of the ring $\RR$ where $E_i:=E(i):=\alpha\circ h(i)$ for some computable (more precisely, $(\nu_\IN,\id_{\s})$-computable) bijection $h:\IN\to \dom(\alpha)$. Obviously the functions $(A,B)\mapsto A\cup B$ and  $(A,B)\mapsto A\setminus B$ are also $(E,E,E)$-computable.

Since $\RR$ is countable, $\Omega=\bigcup \RR$  and the measure $\mu$ is finite on $\RR$, the measure is $\sigma$-finite. Since $\mu(\Omega)=\sup_{R\in\RR}\mu(R)$, $\mu(\Omega)$ is a finite $\rho_<$-computable number or $=\infty$. The measure
$\mu$ is computable on $\RR$, iff $\{ (u,v,w)\mid \nu_\IQ(u)<\mu(\alpha(v))<\nu_\IQ(w)\}$ is r.e.

From the notation $\alpha$ of the ring a representation $\delta$ of the $\sigma$-algebra ${\AA}({\RR})$ can be defined inductively as follows:
\begin{eqnarray*}
\delta(0\langle w\rangle 000\ldots)&:= &\alpha(w)\,,\\
\delta(1p)\hspace{9ex} &:=& \Omega\setminus \delta(p)\,,\\
\delta(2\langle p_0,p_1,\ldots\rangle)&:=&\bigcup_{i\in\IN}\delta(p_i)\,.
\end{eqnarray*}
In this case, if $\delta(p)=B$ then $p$ encodes a finite-path tree (a term) which protocols the generation of the set $B$ from ring elements by repeated application of the unary operation ``complement'' and the $\omega$-ary operation  ``countable union''. The tremendous amount of information contained in a $\delta$-name is not really necessary if we are only interested in computing the measure of the set.
Instead, for given measure $\mu$ the $\sigma$-algebra $\AA$ is factorized by the equivalence relation $A\sim_\mu B \iff \mu(A\syd B)=0$.

\medskip
\noindent {\bf In the following let $\mu$ be a computable measure on the computable $\sigma$-algebra $(\Omega,\AA,\RR,\alpha)$.}
\medskip

We define various representations of the class $[\AA]$. By Lemma~\ref{l6} and Definition~\ref{d1}, $[A]$ is defined uniquely by the set of all $\mu(A\cap E)$ for $E\in \RR$, see Lemma~\ref{l5}.


\begin{defi}\label{d2} Define representations $\zeta_+,\zeta_-$ and 
  $\zeta$ of $\,[\AA]$ as follows:
\begin{enumerate}
\item \label{d2a} $\zeta_+(p)=[A]$ \ \ iff $p$ is (encodes) a list of
  all $\langle u,v\rangle$ such that
\[\nu_\IQ(u)<\mu(\alpha(v)\cap A),\]
\item \label{d2b} $\zeta_-(p)=[A]$ \ \ iff $p$ is (encodes) a list of 
  all $\langle v,w\rangle$  such that 
\[\mu( \alpha(v)\cap A)<\nu_\IQ(w),\]
\item\label{d2c} $\zeta(p)=[A]\phantom{_+}$ \ \ iff $p$ is (encodes) a list of 
  all $\langle u,v,w\rangle$  such that 
\[\nu_\IQ(u)< \mu(\alpha(v)\cap A)<\nu_\IQ(w).\]
\end{enumerate}
\end{defi}\smallskip

\noindent A $\zeta_+$-name of a set $A$ consists of all rational
lower bounds of the $\mu(R\cap A)$ ($A\in\RR$). Since the numbers
$\mu(R)$ are $\rho$-computable, a $\zeta_-$-name of $A$, yields a list
of all rational lower bounds of $\mu(R\setminus A)$ ($A\in\RR$)
(Definition~\ref{d10}, Lemma~\ref{l12}).  In \cite{WD06} rational
lower bounds of $\mu(A\setminus R)$ instead of $\mu(R\setminus A)$ are
used for defining representations which then differ significantly from
the ones defined here.

We must show that the definitions do not depend on the representative $A$ of the class~$[A]$.

\begin{lem}\label{l5}
The representations in Definition~\ref{d2} are well-defined.
\end{lem}

\proof 
Suppose  $\zeta_+(p)=[A]$ and  $\zeta_+(p)=[B]$ according to Definition~\ref{d2}. Then for all $r\in \IQ$ and for all $R\in \RR$,
\begin{eqnarray}
\label{f12}r<\mu(R\cap A) &\iff & r<\mu(R\cap B)\,,
\end{eqnarray}
hence $\mu(R\cap A)=\mu(R\cap B)$ for all $R\in \RR$. By Lemma \ref{l6} $\mu(A\syd B)=0$ and hence $[A]=[B]$.

The argument is the same for $\zeta$. For the case $\zeta_-$ replace ``$<$'' in (\ref{f12}) by ``$>$''.
\qq

The representation $\zeta_+$ ($\zeta_-, \ \zeta$) is the poorest representation that allows to compute $\rho_<$-names ($\rho_>$-names, $\rho$-names) of all $\mu(\alpha(v)\cap A)$.

\begin{lem}\label{l8} For every representation $\gamma$ of a subset of $\,[\AA]$,
\begin{enumerate}
\item \label{l8a}$\gamma\leq \zeta_+  \ \iff \  ([A],R)\mapsto \mu(R\cap A)\ \ \mbox{is}\ \ (\gamma,\alpha,\rho_<)\mbox{-computable}$,
\item \label{l8b}$\gamma\leq \zeta_-  \  \iff \   ([A],R)\mapsto \mu(R\cap A)\ \ \mbox{is}\ \ (\gamma,\alpha,\rho_>)\mbox{-computable}$,
\item \label{l8c}$\gamma\leq \zeta \ \ \iff \   ([A],R)\mapsto \mu(R\cap A) \ \ \mbox{is}\ \ (\gamma,\alpha,\rho)\mbox{-computable}$.
\end{enumerate}
\end{lem}

\proof
The statements can be derived from a general theorem \cite[Theorem~13.1]{WG09}.
We give a direct proof here.

(\ref{l8a})  There is a Type-2 machine $M$ that on input $(p,v)\in\om\times \s$ computes a list of all $u\in\s$ such that $(u,v)$ is listed in $p$\,. If $\zeta_+(p)=[A]$ and $\alpha(v)=R$ then
$f_M(p,v)$ is a list of all $u$ such that $\nu_\IQ(u)< \mu(R\cap A)$, hence
$\rho_<\circ f_M(p,v)=\mu(R\cap A)$. Therefore, the function $([A],R)\mapsto \mu(R\cap A)$ is $(\zeta_+,\alpha,\rho_<)$-computable. Consequently, this function is
$(\gamma,\alpha,\rho_<)$-computable if $\gamma\leq \zeta_+$.

On the other hand, suppose that the function $([A],R)\mapsto\mu(R\cap A)$ is
$(\gamma,\alpha,\rho_<)$-computable. Then there is a Type-2 machine $M$ which on input $(p,v)\in\dom(\gamma)\times \dom(\alpha)$ writes a list  of all $u\in\dom(\nu_\IQ)$, such that $\nu_\IQ(u)<\mu(\alpha(v)\cap\gamma(p))$. From $M$ we can construct a Type-2 machine $N$ which on input $p$ writes a list of all $(u,v)\in\dom(\nu_\IQ)\times \dom(\alpha)$ such that the machine $M$ on input $(p,v)$ writes $u$ in finitely many steps of computation.
Therefore, $f_N(p)$ is a list of all $(u,v)$ such that $\nu_\IQ(u)<\mu(\alpha(v)\cap\gamma(p))$ hence $\zeta_+\circ f_N(p)=\gamma(p)$. We obtain $\gamma\leq \zeta_+$.

(\ref{l8b}) and (\ref{l8c}) can be proved accordingly.
\qq

Therefore, $\zeta_+$ is (up to equivalence) the poorest representation $\gamma$ of $[\AA]$ such that $([A],R)\mapsto \mu(R\cap A)$ is $(\gamma,\alpha,\rho_<)$-computable etc.

For representations $\gamma$ and $\delta$, $\gamma\wedge \delta$ is the greatest lower bound of $\gamma$ and $\delta$ for the reducibility $\leq$, where $(\gamma \wedge \delta)\langle p,q\rangle =x\iff \gamma(p)=\delta(q)=x$
\cite[Section~3.3]{Wei00}. Remember that for the well-known representations of the real numbers,
$\rho\equiv \rho_<\wedge\rho_>$ \cite[Lemma~4.1.9]{Wei00}.

\begin{lem}\label{l1} $ $\hfill
\begin{enumerate}
\item \label{l1a} $\zeta\equiv \zeta_+\wedge\zeta_-$, in particular, $\zeta\leq \zeta_+$, \ \ $\zeta\leq \zeta_-$ and $\zeta_+\wedge\zeta_-\leq \zeta$.
\item \label{l1b} The function $[A]\mapsto [A^c]$ is $(\zeta_+,\zeta_-)$-computable and $(\zeta_-,\zeta_+)$-computable.
\end{enumerate}
\end{lem}

\proof

(\ref{l1a}) By Lemma~\ref{l8}(\ref{l8c}), the function  $h: ([A],R)\mapsto \mu(R\cap A)$ is $(\zeta,\alpha,\rho)$-computable. Since
 $\rho\leq \rho_<$, the function $h$ is $(\zeta,\alpha,\rho_<)$-computable, hence
$\zeta\leq\zeta_+$ by Lemma~\ref{l8}(\ref{l8a}). Accordingly, $\zeta\leq\zeta_-$. Therefore, $\zeta\leq \zeta_+\wedge \zeta_-$.

On the other hand, since $\zeta_+\wedge \zeta_-\leq \zeta_+$, by Lemma~\ref{l8}(\ref{l8a}) the function $h$ is $(\zeta_+\wedge \zeta_-,\alpha,\rho_<)$-computable. Accordingly, the function $h$ is $(\zeta_+\wedge \zeta_-,\alpha,\rho_>)$-computable. Since $\rho_<\wedge\rho_>\leq\rho$, the function $h$ is $(\zeta_+\wedge \zeta_-,\alpha,\rho)$-computable. Finally,
$\zeta_+\wedge \zeta_-\leq \zeta$ by Lemma~\ref{l8}(\ref{l8c}).

(\ref{l1b})  Let $c[A]:=[\Omega\setminus A]=[A^c]$. By Lemma~\ref{l8}(\ref{l8a}) the function $G:([A], R)\mapsto \mu(R\cap A)$ is $(\zeta_+,\alpha,\rho_<)$-computable. Since $R\mapsto \mu(R)$ is $(\alpha,\rho)$-computable, the function
$([A], R)\mapsto \mu(R) - \mu(R\cap A)$ 
is $(\zeta_+,\alpha,\rho_>)$-computable, hence
$\mu(\alpha(w))-\mu(\alpha(w)\cap \zeta_+(p))=\rho_>\circ d(p,w)$ for some computable function $d$. We obtain
$G(c\circ \zeta_+(p),\alpha(w))=\mu(\alpha(w)\cap c\circ\zeta_+(p))=
\mu(\alpha(w))-\mu(\alpha(w)\cap \zeta_+(p))=\rho_>\circ d(p,w)$.
By Lemma~\ref{l8}(\ref{l8b}), $c\circ \zeta_+\leq \zeta_-$, hence $c$ is $(\zeta_+,\zeta_-)$-computable.

$(\zeta_-,\zeta_+)$-computability of complementation can be proved accordingly.
\qq

For the representations $\zeta_+$, $\zeta_-$ and $\zeta$ a name of a class $[A]$
allows to compute $\mu(\alpha(v)\cap A)$ w.r.t. $\rho_<$, $\rho_>$ and $\rho$, respectively. Since $(\alpha(v)\cap A)\cup (\alpha(v)\setminus A)=\alpha(v)$ and a $\rho$-name of $\mu(\alpha(v))$ is known for the computable measure space, from a $\rho_<$-name
($\rho_>$-name, $\rho$-name) of $\mu(\alpha(v)\cap A)$ we can compute a $\rho_>$-name ($\rho_<$-name, $\rho$-name) of  $\mu(\alpha(v)\setminus A)$ and vice versa. Therefore, we can define representations such that names allow to compute all $\mu(\alpha(v)\setminus A)$ which are equivalent to the former ones.

\begin{defi}\label{d10}$ $
 \begin{enumerate}
\item\label{d10a} $\zeta'_+(p)=[A]$  \ \ iff $p$ is (encodes) a list of 
  all $(u,v)$ such that 
\[\mu(\alpha(v)\setminus A)<\nu_\IQ(u),\]
\item\label{d10b} $\zeta'_-(p)=[A]$  \ \ iff $p$ is (encodes) a list of 
  all $(u,v)$ such that 
\[\nu_\IQ(u)<\mu(\alpha(v)\setminus A).\]
\item\label{d10c} $\zeta'(p)=[A]\phantom{_+}$\ \ iff $p$ is (encodes) a list of 
  all $(u,v,w)$  such that
\[\nu_\IQ(u)< \mu(\alpha(v)\setminus A)<\nu_\IQ(w).\]
\end{enumerate}
\end{defi}

\noindent Notice that
\begin {eqnarray}\label{f17}
\zeta_-'(p)=[A]\iff \zeta_+(p)=[A^c] & \mbox{and}& \zeta_+'(p)=[A]\iff \zeta_-(p)=[A^c]\,.
\end{eqnarray}

\begin{lem}\label{l12} $\zeta_+\equiv \zeta_+'$, $\zeta_-\equiv \zeta_-'$,
$\zeta\equiv \zeta'$
\end{lem}

\proof Straightforward.
\qq

There is a computable measure on a computable $\sigma$-algebra such that
$\zeta_+\not\leq\zeta$  (see the proof of
Theorem~\ref{t3} (\ref{t3b}) below). As usual already translation by a continuous function is impossible, $\zeta_+\not\leq_t\zeta$.
We determine the degree of unsolvability of the translations from $\zeta_+$ to $\zeta$ and the other similar ones.

Let $\XX_1,\YY_1,\XX_2,\YY_2$ be represented sets and let $f_1:\XX_1\mto \YY_1$ and $f_2:\XX_2\mto\YY_2$ be multifunctions. Then $f_1\leq_W f_2$ iff there are computable functions $G,H$ on $\om$ such that for all realizations $F_2$ of $f_2$, \
$F_1:p\mapsto H(p,F_2\circ G(p))$ realizes $f_1$ (\cite{BG11,BG11a}, where $\leq_W$ is called Weihrauch reducibility). This means that composition with $G$ and $H$ in this manner transforms every realization of $f_2$ to a realization of $f_1$. The multi-functions $f_1,f_2$ are called W-equivalent, $f_1\equiv_W f_2$,  iff $f_1\leq_W f_2$ and $f_2\leq_W f_1$. A stronger reducibility is defined by
$f_1\leq_{sW} f_2$ iff there are computable functions $G,H$ on $\om$ such that for all realizations  $F_2$ of $f_2$, \ $F_1:p\mapsto H\circ F_2\circ G(p)$ realizes $f_1$ \cite{BG11,BG11a}. Obviously, $f_1\leq_{sW} f_2$ implies $f_1\leq_W f_2\,.$

It is known  that $\rho_n\not\leq \rho_c$, $\rho_<\not\leq \rho_c$, $\rho_>\not\leq \rho_c$, $\rho_<\not\leq \rho_>$ and $\rho_>\not\leq \rho_<$ ,  where $\rho_n$ is the representation of the real numbers by
(not necessarily fast) converging sequences of rational numbers \cite{Wei00}.
These five translation problems are of the same sW-degree of unsolvability.
Furthermore, the identity ${\rm ECf}:(2^\IN,{\rm En})\to (2^\IN, {\rm Cf})$ and
complementation of enumeration ${\rm CE}:(2^\IN,{\rm En})\to (2^\IN, {\rm En})$,
$K\mapsto \IN\setminus K$, are in this sW-degree \cite{Wei92a}, where ${\rm Cf}$ is the characteristic function representation and
${\rm En}$ is the enumeration representation  of the subsets of $\IN$ \cite{Wei00}.

Let ${\rm En}^*:\om\to 2^{\s}$ be the canonical enumeration representation of the set of subsets of $\s$, that is, $\iota(w)$ is a subword of $p\in\om$ iff $w\in {\rm En}^*(p)$.
Then also complementation ${\rm CE}^*:(2^\s,{\rm En}^*)\to (2^\s,{\rm En}^*)$, $W\mapsto \s\setminus W$, is in the  sW-degree of $\rm CE$. Finally, it is known that
\begin{eqnarray}\label{f13}
f\leq_W{\rm CE} \iff f\leq_{sW}{\rm CE}\ \ & \ \mbox{for all functions $f$ on represented sets}.
\end{eqnarray}

\begin{thm}\label{t3} For a given computable measure on a computable $\sigma$-algebra $\AA$ define $\AA_+:=([\AA],\zeta_+)$, \ \ $\AA_-:=([\AA],\zeta_-)$, \ \ $\AA_0:=([\AA],\zeta)$,
and for $a,b\in\{+,-,0\}$ define the translation
$\id_{ab}:\AA_a\to\AA_b $ by $  \id_{ab}([A]):=[A]$. Then:
\begin{enumerate}
\item \label{t3a}For every computable measure on a computable $\sigma$-algebra,
 \begin{eqnarray}
 \label{f15}&\id_{+-}\leq_{sW} \id_{+0}\leq_{sW} {\rm CE}\,, \  &  \id_{+0}\leq_W \id_{+-}\,,\\
\label{f14}&\id_{-+}\leq_{sW} \id_{-0}\leq_{sW} {\rm CE}\,, \  &  \id_{-0}\leq_W \id_{-+}\,.
\end{eqnarray}
 \item \label{t3b}There is a computable probability measure on a computable $\sigma$-algebra such that
\[ \id_{+-}\ \equiv_{sW} \ \id_{+0}\ \equiv_{sW} \ \id_{-+}\ \equiv_{sW} \ \id_{-0}\ \equiv_{sW}\ {\rm CE}\,.\]
\end{enumerate}
\end{thm}

\proof

(\ref{t3a})
We prove \,$\id_{+-}\leq_{sW}\id_{+0}$. By Lemma~\ref{l1} there is a computable function $H$ such that $\zeta(q)=\zeta_-\circ H(p)$.  Define $G(p):=p$. Suppose $F_2$ realizes $\id_{+0}$, that is $\zeta_+(p)=\zeta\circ F_2(p)$.
Then
 \[ \zeta_+(p)=\zeta\circ F_2(p) =\zeta_-\circ H \circ F_2\circ G(p) \,,  \]
 hence $p\mapsto H\circ F_2\circ G(p)$ realizes $\id_{+-}$.
 Since $H$ and $G$ are computable, $\id_{+-}\leq_{sW}\id_{+0}$.
 \medskip

We prove \,$\id_{+0}\leq_W\id_{+-}$. By Lemma~\ref{l1} there is a computable function $h$ such that $(\zeta_+\wedge \zeta_-)(q)=\zeta\circ h(p)$. Define $H$ and $G$ by $H(p,q):=h(\langle p,q\rangle))$ and $G(p):=p$. Suppose $F_2$ realizes $\id_{+-}$, that is $\zeta_+(p)=\zeta_-\circ F_2(p)$. Then
\[\zeta_+(p)=\zeta_-\circ F_2(p)=(\zeta_+\wedge \zeta_-)\langle p, F_2(p)\rangle =
\zeta\circ h \langle p, F_2(p)\rangle=\zeta\circ H(p,F_2\circ G(p)) \,,   \]
 hence $p\mapsto H(p,F_2\circ G(p))$ realizes $\id_{+0}$.
 Since $H$ and $G$ are computable, $\id_{+0}\leq_W\id_{+-}$.
\medskip

We prove ${\rm id}_{+-}\leq_{sW} {\rm CE}$. By Definition~\ref{d2}, $\zeta_+(p)=[A]$ means: \\ for all $x\in\s$,
$\iota(x)$ is a subword of  $p$ iff
\[(\exists u\in\dom(\nu_\IQ))(\exists v\in\dom(\alpha))\,(x=\langle u,v\rangle \an \nu_\IQ(u)<\mu(\alpha(v)\cap A))\,.\]
There is a Type-2 machine $M$ that on input $q\in\om$ writes a list of all $\langle v,w\rangle$ such that $v\in\dom(\alpha)$ and for some $u\in\dom(\nu_\IQ)$,   $\langle u,v\rangle$ is listed in $q$ and $\nu_\IQ(w)>\nu_\IQ(u)$.

Let $F_2$ be a realization of ${\rm CE}^*$. Suppose $\zeta_+(p)=[A]$. Let $q:= F_2(p)$. By the definition of $M$, $f_M(q)$ is a list of words $\langle v,w\rangle$ such that $v\in\dom(\alpha)$ and $w\in\dom(\nu_\IQ)$.
Suppose $\mu(\alpha(v)\cap A)<\nu_\IQ(w)$. Then for some $u$,
$\mu(\alpha(v)\cap A)\leq \nu_\IQ(u) <\nu_\IQ(w)$.
By the definition of $\zeta_+$,  $\langle u,v\rangle$ is not listed in $p$ and hence listed in $F_2(p)$.
Therefore, $\langle v,w\rangle$ is listed in $f_M\circ F_2(p)$.
On the other hand suppose $\langle v,w\rangle$ is listed in $f_M\circ F_2(p)$. Then for some $u$, $\langle u,v\rangle$ is listed in $F_2(p)$ and $ \nu_\IQ(u) <\nu_\IQ(w)$. Therefore, $\langle u,v\rangle$ is not listed in $p$, hence
$\mu(\alpha(v)\cap A)\leq \nu_\IQ(u)<\nu_\IQ(w)$.

Combining the two cases we obtain, $\zeta_+(p)=\zeta_-\circ f_M\circ F_2(p)$. Therefore the function $f_M\circ F_2$ realizes the function ${\rm id}_{+-}$\;, hence  ${\rm id}_{+-}\leq_{sW} {\rm CE}^*\leq_{sW} {\rm CE}$.

In summary $\id_{+-}\leq_{sW}\id_{+0}\leq_W\id_{+-}\leq_{sW} {\rm CE}$. Applying (\ref{f13}) we obtain (\ref{f15}).
\medskip

(\ref{f14}) can be proved accordingly.

\medskip
(\ref{t3b})
Let $\Omega:=\IN$, $\AA:=2^\IN$, $\RR:=$ the set of finite subsets of $\IN$ with canonical notation $\alpha $, and for $A\in\AA$ let $\mu(A):=\sum _{i\in A}3^{-i}$. Then $\mu$ is a computable measure on the computable $\sigma$-algebra $(\Omega,\AA,\RR,\alpha)$ such that $\mu(\Omega)=3/2$.

First, we prove ${\rm CE}\leq_{sW}\id_{+-}\,$.

We show that the function $h:2^\IN\to [\AA]$, $h(A):=[A]$ is $({\rm En},\zeta_+)$-computable. Suppose ${\rm En}(p)=A$.
There is a Type-2 machine $M$ that on input $p\in\dom({\rm En})$ produces a list of all $\langle u,v\rangle$ such that
$\nu_\IQ(u)<\mu(\alpha(v)\cap A_k)$ for some $k$, where $A_k$ is the set of all $n$ such that $01^{n+1}0$ is a subword of the first $k$ symbols of $p$. If $\langle u,v\rangle $ is in this list then $\nu_\IQ(u)<\mu(\alpha(v)\cap A_k)\leq \mu(\alpha(v)\cap {\rm En}(p))$. If
$\nu_\IQ(u)< \mu(\alpha(v)\cap {\rm En}(p))$, then $\nu_\IQ(u)<\mu(\alpha(v)\cap A_k)$ for some $k$, hence $\langle u,v\rangle$ is in the list. Therefore, $\zeta_+\circ f_M(p)=[{\rm En}(p)]$, hence $f_M$ realizes~$h$.

We show that the function $h':[\AA]\to 2^\IN$, $h'([A]):= \IN\setminus A$ is $(\zeta_-,{\rm En})$-computable. Since $[A]=[B]$ implies $A=B$, the function $h'$ is well-defined.
There is a computable function $g:\IN\to\s$ such that $(\forall i)\,\alpha\circ g(i)=\{i\}$. And there is a computable function $d:\IN\to\s$ such that $\nu_\IQ\circ d(i)=3^{-i}$. There is a Type-2 machine $N$ that on input $q\in\dom(\zeta_-)$ lists all $01^{i+1}0$ such that $(g(i),d(i))$ is listed in $q$. Suppose, $[A]=\zeta_-(q)$. Then $q$ is a list of all $(v,w)$ such that $\mu(A\cap\alpha(v))<\nu_\IQ(w)$. Since
\begin{eqnarray*}
i\not\in A &\iff& \mu(A\cap \alpha\circ g(i)) < \nu_\IQ\circ d(i)\\
&\iff& \mbox{$(g(i),d(i))$ is listed in $q$}\iff \mbox{$01^{i+1}0$ is a subword of $f_N(q)$}\,,
\end{eqnarray*}
$h'\circ \zeta_-(q)=h'([A])= \IN\setminus A={\rm En}\circ f_N(q)$.
Therefore, $f_N$ realizes $h'$.

Suppose $F_2$ realizes $\id_{+-}$. Then $f_N\circ  F_2\circ f_M$ realizes
$h'\circ {\rm id}_{+-}\circ h ={\rm CE}$. Therefore, ${\rm CE}\leq_{sW}\id_{+-}$.
By (\ref{f15}), ${\rm CE}\leq_{sW} \id_{+-}\leq_{sW} \id_{+0}\leq_{sW} {\rm CE}$,
hence $\id_{+-}\equiv_{sW} \id_{+0}\equiv_{sW} {\rm CE}$.

$\id_{-+}\equiv_{sW} \id_{-0}\equiv_{sW} {\rm CE}$ can be proved accordingly.
\medskip

Let $\mu':= 2/3\cdot\mu$. Then $\mu'(\Omega)=1$, hence $\mu'$  is a probability measure and the results hold as well for $\mu'$.
\qq

\begin{lem}\label{l7}
The function $([A],G)\mapsto [A\cap G]$ for $A\in \AA$  and $G\in\RR$ is
$(\zeta_+,\alpha,\zeta_+)$-computable, $(\zeta_-,\alpha,\zeta_-)$-computable and $(\zeta,\alpha,\zeta)$-computable.
\end{lem}

\proof

$(\zeta_-,\alpha,\zeta_-)$: There is a computable word function $h$  such that
$\alpha(v)\cap\alpha(u)=\alpha\circ h(v,u)$ (see the remark after Definition~\ref{d1}).

Suppose  $\zeta_-(p)=[A]$, $\alpha(u)=G$ and $\zeta_-(q)=[A\cap G]$. Then
$p$ is a list of all $(v,w)$ such that $\mu(\alpha(v)\cap A)<\nu_\IQ(w)$. Correspondingly,
$q$ is a list of all $(v,w)$ such that $\mu(\alpha(v)\cap G\cap A)<\nu_\IQ(w)$, that is, $q$ is a list of all $(v,w)$ such that $\mu(\alpha\circ h(v,u)\cap A)<\nu_\IQ(w)$, hence $q$ is a list of all $(v,w)$ such that $(h(v,u),w)$ is in the list $p$.
There is a machine that on input $(p,u)$ writes a list of all
$(v,w)$ such that $( h(v,u),w)$ is listed in $p$. Therefore, $([A],G)\mapsto [A\cap G]$ is $(\zeta_-,\alpha,\zeta_-)$-computable.

The other two statements can be proved accordingly.
\qq

Let $\overline\rho_<$ and $\overline \rho_>$  be the lower and upper representation of $\overline\IR:=\IR\cup\{-\infty,\infty\}$, respectively, and let $\overline \rho=\overline\rho_<\wedge \overline \rho_>$ \cite[Secton~4.1]{Wei00}.  Informally, $\overline\rho_<(p)=x$ iff $p$ is a list of all $a\in\IQ$ such that $a<x$, and
$\overline\rho_>(p)=x$ iff $p$ is a list of all $a\in\IQ$ such that $a>x$.

\begin{lem}\label{l10} $ $
\begin{enumerate}
\item \label{l10a} $\mu:[A]\mapsto \mu(A)$  is $(\zeta_+,\overline\rho_<)$-computable,
\item \label{l10b} $\mu:[A]\mapsto \mu(A^c)$  is $(\zeta_-,\overline\rho_<)$-computable,
\item \label{l10d} $\mu(\Omega)$ is $\overline\rho_<$-computable,
\item \label{l10c} For finite measures, $\mu:[A]\mapsto \mu(A)$ is $(\zeta,\rho)$-computable iff $\mu(\Omega)$ is $\rho$-computable.
\end{enumerate}
\end{lem}

\proof

(\ref{l10a}) Since $\Omega=\bigcup \RR$ and $\RR$ is countable and closed under union, $\mu(A)=\mu(A\cap\Omega)=\sup_{R\in\RR}\mu( R\cap A)$.
There is a Type-2 machine $M$ which on input $p$ writes a list $q$ of all $u$ such that for some $v$, $(u,v)$ is listed in $p$. If $\zeta_+(p)=[A]$ then $q$ is a list of all $u\in\dom(\nu_\IQ)$ such that $\nu_\IQ(u)<\mu(A)$. Therefore  $f_M$ is a  $(\zeta_+,\overline\rho_<)$-realization of
$[A]\mapsto \mu(A)$.
\medskip

(\ref{l10b}) Suppose $\zeta_-'(p)=[A]$. By (\ref{f17}), $\zeta_+(p)=[A^c]$, hence by (\ref{l10a}), $\mu(A^c)=\overline\rho_<\circ f(p)$ for some computable function $f$. Therefore $[A]\mapsto \mu(A^c)$ is $(\zeta_-',\overline\rho_<)$-computable and hence $(\zeta_-,\overline\rho_<)$-computable by Lemma~\ref{l12}.
\medskip

(\ref{l10d}) This follows from (\ref{l10a}) since  $[\Omega]$ is $\zeta_+$-computable.
\medskip

(\ref{l10c}) Suppose $\mu(\Omega)$ is $\rho$-computable. Since by Lemma~\ref{l1} $\zeta\leq \zeta_+$ and $\zeta\leq\zeta_-$,
$[A]\mapsto \mu(A)$ and  $[A]\mapsto \mu(A^c)$ are $(\zeta,\overline \rho_<)$-computable by (\ref{l10a}) and (\ref{l10b}) above. Since $\mu(A)+\mu(A^c)=\mu(\Omega)$ and $\mu(\Omega)$ is $\rho$-computable, $[A]\mapsto \mu(A)$ is $(\zeta,\rho)$-computable.

Suppose $[A]\mapsto \mu(A)$ is $(\zeta,\rho)$-computable. Since $[\Omega]=\zeta(p)$ for some computable $p\in\om$, $\mu(\Omega)$ must be $\rho$-computable.
\qq

\begin{exa}[non-computable $\mu(\Omega)$] \label{ex1}{\rm
 Let $\Omega:=\IN$, $\AA:=2^\IN$, $\RR:=$ the set of finite subsets of $\IN$ with canonical notation $\alpha$ and $\mu(A):=\sum _{i\in A}2^{-h(i)}$ where $h:\IN\to\IN$ is an injective computable numbering of some r.e. set $K\In\IN$ that is not recursive. Then $\mu$ is a computable measure on the computable $\sigma$-algebra
$(\Omega,\AA,\RR,\alpha)$. There is a computable $p\in\om$ such that $\zeta(p)=[\Omega]=[\IN]$.
Since $\mu(\Omega)=\sum_{i\in\IN}2^{-h(i)}=\sum_{n\in K}2^{-n}$,
$\mu(\Omega)$ is $\rho_<$- computable but not $\rho$-computable \cite[Example~4.2.4]{Wei00}. \qq
}\end{exa}

\section{Representations of the sets of finite measure}\label{sece}

In this section we introduce and study representations of the set $[\AA^f])$ for the set $\AA^f$ of measurable sets of finite measure. $\mu(\Omega)$ may be finite or infinite.
By Theorem~\ref{t2}, $(\AA^f,d)$ with $d(A,B)=\mu(A\syd B)$ is a complete pseudometric  space with $\RR^f$ as a dense subset. Remember that for our computable measure $\mu$ on the computable $\sigma$-algebra,  $\RR^f=\RR$.
Then  $d([A],[B]):=d(A,B)=\mu(A\syd B)$  defines a metric  on the equivalence classes $[\AA^f]=\{[A]\mid A\in \AA^f\}$. (As usual, we use the same symbol $d$ for the pseudometric and its factorization.)

A computable metric space is a quadruple $(M,d,A,\nu)$ such that $(M,d)$ is a metric space, $A\In M$ is dense and $\nu\pf \s\to A$ is a notation of $A$ such that $\dom(\nu)$ is recursive and
the metric $d$ restricted to $A$ is $(\nu,\nu,\rho)$-computable (equivalently, the set of all $(t,u,v,w)$ such that $\nu_\IQ(t)<d(\nu(u),\nu(v))<\nu_\IQ(w)$ is r.e.).
The Cauchy representation of a computable metric space is defined by $\delta_C(p)=x$ iff $p$ is (encodes ) a sequence $v_0,v_1,\ldots \in \s$ such that $d(\nu(v_i),\nu (v_j))\leq 2^{-i}$ if $i<j$ and $x=\lim \nu(v_i)$
\cite[Section~8.1]{Wei00} \cite{BP03}. Notice that $d(x,\nu(v_i))\leq 2^{-i}$.
The metric $d:M\times M\to\IR$ is $(\delta_C,\delta_C,\rho)$-computable.

\begin{lem}\label{l2}
Let $\overline{\AA^f}:=([\AA^f],\ d,\ [\RR],\ \beta)$ where $d([A],[B]):=\mu(A\syd B)$ and $\beta(u):=[\alpha(u)]$.

 \begin{enumerate}
 \item
 $\overline{\AA^f}$ is a complete computable metric space.
  \item For the Cauchy representation $\xi_C$ of $\overline{\AA^f}$ the measure $\mu:[A]\mapsto\mu([A])=\mu(A)$ is $(\xi_C,\rho)$-computable.
 \end{enumerate}
\end{lem}

\proof
By Theorem~\ref{t2}, $(\AA^f,d)$ with $d(A,B)=\mu(A\syd B)$ is a complete pseudometric  space with $\RR^f= \RR$ as a dense set. Since by Definition~\ref{d6}, $d([A],[B])=0 \iff [A]=[B]$,
$([\AA^f],\ d)$ is a complete metric space with $[\RR]$ as a dense subset. Obviously $\beta$ is a notation of $[\RR]$ with recursive domain.
Since $d([A],[B])=\mu(A\syd B)$ and by Definition~\ref{d1}, the symmetric difference on $\RR$ is computable and $\mu$ is $(\alpha,\rho)$-computable, the metric on
$[\RR]$ is $(\beta,\beta,\rho)$-computable. Therefore, $\overline{\AA^f}$ is a computable metric space.

Since $d$ is $(\xi_C,\xi_C,\rho)$-computable, $\xi_C(q)=[\emptyset]$ for some computable $q\in\om$ and $\mu(A)=\mu(\emptyset\syd A)=d([\emptyset],[ A])$, the measure $\mu$ is $(\xi_C,\rho)$-computable.
\qq

We introduce two further representations of the set $[\AA^f]$ of measurable sets of finite measure by adding the measure of $A$ to the  $\zeta$-names of $[A]\in [\AA^f]$.

\begin{defi}\label{d7}
For the space $\overline{\AA^f}:=([\AA^f],\ d,\ [\RR],\ \beta)$ let $\xi_C$ be the Cauchy representation and define representations $\xi_+,\xi_-$ and $\xi$ by
\begin{eqnarray*}
\xi_+\langle p,q\rangle =[A] &:\iff &\zeta_+(p)=[A] \ \mbox{and}\ \ \rho_>(q)=\mu(A)\,,\\
\xi_-\langle p,q\rangle =[A] &:\iff &\zeta_-(p)=[A] \ \mbox{and}\ \ \rho_<(q)=\mu(A)\,,\\
\xi\langle p,q\rangle =[A] &:\iff &\ \ \zeta(p)=[A] \ \mbox{and}\ \ \rho(q)=\mu(A)\,.
\end{eqnarray*}
\end{defi}

\begin{thm}\label{t4} On the space $\overline{\AA^f}$,
\begin{enumerate}
\item \label{t4c}
$[A]\mapsto\mu(A)$ is $(\xi_+, \rho_>)$-computable, $(\xi_-, \rho_<)$-computable and $(\xi,\rho)$-computable.
\item \label{t4d} $\xi_+\leq \zeta_+$, $\xi_-\leq \zeta_-$ and $\xi\leq \zeta$,
\item \label{t4a} $\xi_C\equiv \xi_+\equiv \xi_-\equiv \xi$,
\item\label{t4b}
$\xi\equiv \zeta$ iff $\mu(\Omega)$ is $\rho$-computable.
\end{enumerate}
\end{thm}

\proof
(\ref{t4c}), (\ref{t4d}) Obvious.
\smallskip


(\ref{t4a})
$\xi\leq \xi_+$\,: Form a $\xi$-name of $[A]$ we can compute a $\zeta$-name of
$[A]$ and a $\rho$-name of $\mu(A)$.
Since $\zeta\leq\zeta_+$ and $\rho\leq \rho_>$ we can compute a $\zeta_+$-name $p'$ of $[A]$ and a $\rho_>$-name $q'$ of $\mu(A)$. Then $\langle p',q'\rangle $ is a $\xi_+$-name of $[A]$.
\smallskip

$\xi\leq \xi_-$\,: Accordingly.

\smallskip
$\xi_+\leq \xi_C$\,: Since $\RR$ is dense in $\AA^f$ (Theorem~\ref{t2}) for every $A\in\AA^f$ and
$\varepsilon >0$ there is some $R\in \RR$ such that $\mu(A \syd R)<\varepsilon$.  
Notice that
\[\mu(A\syd R)=\mu(A\setminus R)+\mu(R\setminus A)=\mu(A)-\mu(A\cap R) +\mu(R)-\mu(A\cap R)\,.\]
The function $[A]\mapsto \mu(A)$ is $(\xi_+,\rho_>)$-computable by Definition~\ref{d7}, the function
$R\mapsto\mu(R)$ is $(\alpha,\rho_>)$-computable,
and by $\xi_+\leq\zeta_+$ and Lemma~\ref{l8}(\ref{l8a}) the function
$([A],R)\mapsto -\mu(A\cap R)$ is $(\xi_+,\alpha,\rho_>)$-computable. Therefore,
$G: ([A], R)\mapsto \mu(A\syd R)$ is $(\xi_+,\alpha,\rho_>)$-computable.
There is a machine $M$ which on input $(p,v)$ writes a sequence of all (codes of)
$a\in\IQ$ such that $G(\xi_+(p),\alpha(v))< a$.

There is a machine $N$ which on input $p$ writes a sequence $v_0,v_1,\ldots$ of words where $v_i$ is computed as follows: $N$ runs $M$ as a subprogram and searches some $(v_i,k)$ such that $M$ on input $(p,v_i)$ writes the rational number $2^{-i-1}$ in at most $k$ steps of computation. If $\xi(p)=[A]$ then for every $i$ the search for $v_i$ is successful. Since $\mu(\xi(p)\syd \alpha(v_i))< 2^{-i-1}$, $\xi_C(v_0,v_1,\ldots)=[A]$. Therefore, $\xi_+\leq \xi_C$. \\
 $\xi_-\leq \xi_C$ can be proved accordingly..
\medskip

$\xi_C\leq \xi$\,: Suppose $\xi_C(r)=[A]$. Then $r$ is (encodes) a sequence $R_0,R_1,\ldots$ of ring elements such that $d(R_i,A)=\mu(R_i\syd A)\leq 2^{-i}$.
We must compute $\mu(A)$ and furthermore prove $\xi_C\leq \zeta$ (that is, we must compute a $\zeta$-name of $[A]$).

Since $\mu(R_i\syd A)\leq 2^{-i}$, for every $R\in\RR$,
$\mu((R\cap R_i)\syd (R\cap A))=\mu(R\cap (R_i\syd A))\leq 2^{-i}$, hence
\begin{eqnarray}\label{f18}
|\mu(R\cap R_i)- \mu(R\cap A)|&\leq& 2^{-i}\,.
\end{eqnarray}
 by (\ref{f16}).
Since intersection on $\RR$ is $(\alpha,\alpha,\alpha)$-computable, from an $\alpha$-name of $R$ and  $r\in\dom(\xi_C)$ encoding the sequence $R_0,R_1,\ldots$ we can compute  a sequence $s\in\om$ encoding the sequence  $R\cap R_0,R\cap R_1,\ldots$ which, by (\ref{f18}) is a $\xi_C$-name of $[R\cap A]$. By Lemma~\ref{l2} from $s$ we can compute a $\rho$-name of $\mu([R\cap A])$.
Therefore, $([A],R)\mapsto \mu(A\cap R)$ is $(\xi_C,\alpha,\rho)$-computable.
By Lemma~\ref{l8}, $\xi_C\leq \zeta$. Since $[A]\mapsto\mu(A)$ is $(\xi_C,\rho)$-computable by Lemma~\ref{l2},
$\xi_C\leq \xi$.

(\ref{t4b}) By lemma~\ref{l10}, $\xi\equiv \zeta$ iff $[A]\mapsto\mu(A)$ is $(\zeta,\rho)$-computable iff $\mu(\Omega)$ is $\rho$-computable.
\qq

By Lemma~\ref{l10}(\ref{l10d}), $\mu(\Omega)\in \IR^\infty$ is $\overline \rho_<$-computable, hence $\mu(\Omega)$ is the limit of an increasing  computable sequence of rational numbers which may be finite or $\infty$.
By Example~\ref{ex1} there is a computable finite measure with finite non-computable $\mu(\Omega)$.

If $\mu(\Omega)\in \IR$ is a computable real number, then by Lemma~\ref{l10}(\ref{l10c}), $[A]\mapsto \mu(A)$ is $(\zeta,\rho)$-computable and by Theorem~\ref{t4}(\ref{t4b}), $\zeta\equiv \xi\equiv \xi_C$.

If $\mu(\Omega)\in \IR$ is a computable real number and $\mu(\Omega)>0$  then $\mu':=\mu/\mu(\Omega)$ is a probability measure with the same computability properties.\\

\section{Representations by means of a partition}\label{secd}
We still assume that $\mu$ is a computable measure on the computable $\sigma$-algebra $(\Omega ,\AA,\RR,\alpha)$.
As we have mentioned there are ring elements $F_0,F_1,\ldots$ such that
$(\forall i\neq j)\, F_i\cap F_j=\emptyset, \ \ (\forall i) \mu(F_i)<\infty \ \ \mbox{and}\ \  \bigcup_i F_i=\Omega$  (see (\ref{f4})). Such a sequence $(F_i)_{i\in\IN}$ can be computed.
For $i\in\IN$ define $\mu_i(A):=\mu(A\cap F_i)$. Then every $\mu_i$ is a finite measure and $\mu(A)=\sum_{i\in\IN}\mu_i(A)$.

By Lemma~\ref{l6} for every $A\in \AA$, $[A]$ is defined  by the family
$(\mu(A\cap R))_{ R\in\RR}$. The representations $\zeta_+$, $\zeta_-$ and $\zeta$ from Definition~\ref{d2} are defined by means of this family (``a $\zeta_+$-name of $[A]$ is a list of all $\ldots$'' etc.).
Correspondingly, for every $i$ and $A$, $[A\cap F_i]$ is defined by the family $(\mu_i(A\cap  R))_{ R\in\RR}$.
Therefore $[A]$ is  defined also by the family $(\mu(A\cap F_i\cap R))_ {(i\in\IN,\  R\in\RR)}$ which is a subfamily of $(\mu(A\cap R))_{ R\in\RR}$. We introduce representations $\overline \zeta_+$, $\overline \zeta_-$ and  $\overline \zeta$ of $[\AA]$ by means of  this smaller family and compare them with $\zeta_+$, $\zeta_-$ and $\zeta$.

\begin{defi}\label{d9}$ $
A numbering $F:\IN\to \RR$ is a {\em partition for $\alpha$} iff there is a computable function $g:\IN\to \s$ such that $F(i)=\alpha\circ g(i)$ (that is, $F\leq\alpha$) and
\begin{eqnarray}\label{f28}
&(\forall i\neq j)\, F_i\cap F_j=\emptyset, \ \ (\forall i) \mu(F_i)<\infty
\end{eqnarray}
and it is {\em majorising} if there is a computable function $g'\pf \s\to\IN$ such that
\begin{eqnarray}\label{f29}
&(\forall w\in\dom(\alpha))\ \ \alpha(w)\In \bigcup_{i\leq g'(w)} F_i \,.
\end{eqnarray}
\end{defi}

\begin{lem}\label{l11} There is a majorising partition for $\alpha$.
\end{lem}

\proof \ There is a bijective computable function $h:\IN\to \dom(\alpha)$. For  the numbering $E=\alpha\circ h$ define $F(n):= F_n:= E_n\setminus \bigcup_{i<n} E_i$. Then $F$ satisfies (\ref{f28}).
Since union and set difference are $(E,E,E)$-computable, there is some computable function $g_1:\IN\to\IN$ such that
$F_n=E\circ g_1(n)= \alpha\circ h\circ g_1 $.  Then $g:=h\circ g_1$ is computable and $F=\alpha\circ g$.

Let $g':= h^{-1}$. From the definition of $F$, $E_n=\bigcup_{i\leq n}F_i$, hence
$\alpha(w)=E_{g'(w)}=\bigcup _{i\leq g'(w)}F_i$. Then $g'$ is computable and satisfies (\ref{f29}).
\qq

For a given partition for $\alpha$ we introduce three further representations of $[A]$.
\begin{defi}\label{d8} For a fixed partition  $F$ for $\alpha$  define representations \,$\overline \zeta_+$, $\overline \zeta_-$ and  $\overline \zeta$ of \,$[\AA]$ as follows:
\begin{enumerate}
\item $\overline \zeta_+(p)=[A]$ \ \ iff $p$ is (encodes) a list of 
  all $(u,i,v)$ such that
\[\nu_\IQ(u)<\mu(F_i\cap \alpha(v)\cap A),\]
\item $\overline \zeta_-(p)=[A]$ \ \ iff $p$ is (encodes) a list of 
  all $(i,v,w)$ such that 
\[\mu(F_i\cap \alpha(v)\cap A)<\nu_\IQ(w),\]
\item $\overline \zeta(p)=[A]\phantom{_+}$ \ \ iff $p$ is (encodes) a list of 
  all $(u,i,v,w)$ such that
\[\nu_\IQ(u)<\mu(F_i\cap \alpha(v)\cap A)<\nu_\IQ(w).\]
    \end{enumerate}
\end{defi}\smallskip

\noindent The three representations are well-defined (see Definition~\ref{d2} and Lemma~\ref{l5}).

\begin{thm}\label{t9}
$\zeta_+\leq \overline \zeta_+$,  \ $\zeta_-\leq\overline \zeta_-$ \ and  \ $\zeta\leq \overline \zeta$. Furthermore,
$\zeta_+\equiv \overline \zeta_+$,  \ $\zeta_-\equiv \overline \zeta_-$ \ and  \ $\zeta\equiv \overline \zeta$ if $F$ is majorising.
\end{thm}

\proof Since intersection is computable on $\RR$ there is a computable function $d$ such that $\alpha(v)\cap F_i=\alpha\circ d(v,i)$. There is a Type-2 machine $N$ which on input $p\in \dom(\zeta_+)$ enumerates all $(u,i,v)$ such that
$(u,d(v,i))$ is listed by $p$. Then $f_N$ translates $\zeta_+$ to $\overline \zeta_+$, hence $\zeta_+\leq\overline\zeta_+$.

For proving the other direction let $g'$ be the computable function from  (\ref{f29}). Then
\begin{eqnarray*}
&&\nu_\IQ(u)<\mu(\alpha(v)\cap A) \\
&\iff & \nu_\IQ(u)<\sum_i\mu(F_i\cap\alpha(v) \cap A) \\
&\iff &\nu_\IQ(u)<\sum_{i\leq g'(v)}\mu(F_i\cap\alpha(v) \cap A)\\
&\iff& (\exists u_0,\ldots,u_{g'(v)})\\
&&\big(\nu_\IQ(u)< \sum_{i\leq g'(v)}\nu_\IQ(u_i) \an
(\forall i\leq g'(v))\,\nu_\IQ(u_i)<\mu(F_i\cap\alpha(v)\cap A)\big)\,.
\end{eqnarray*}
There is a Type-2 machine $N$ that on input $p\in\dom(\overline \zeta_+)$ enumerates all $(u,v)$ such that for $k:=g'(v)$ there are $u_0,\ldots,u_k$ with
$\nu_\IQ(u)< \sum_{i\leq k}\nu_\IQ(u_i)$ and $(u_i,i,v)$ can be found in the list $p$ for all $0\leq i\leq k$. The function $f_N$ translates $\overline\zeta_+$ to $\zeta_+$, hence $\overline\zeta_+\leq\zeta_+$.

The other statements can be proved accordingly.
\qq

We introduce a metric $\overline d$ on the $\sigma$-algebra $[\AA]$ and prove that its Cauchy representation is equivalent to $\overline \zeta$. This metric is similar to the metric $d_1$ in \cite[Section 5]{WD05}. We discuss their relation in Section~\ref{secg}  below.

\begin{thm}\label{t5} Let $F$ be a partition for $\alpha$. Then $([\AA],\overline d,[\RR],\beta)$ where $\beta(w):=[\alpha(w)]$ and
\[\overline d([A],[B]):= \overline d(A,B):= \sum _{i\in \IN}\frac{\mu(F_i\cap(A\syd B))}{1+\mu(F_i\cap(A\syd B))}\cdot 2^{-i}\,\]
is a computable metric space such that $\overline\zeta \equiv \overline \xi_C$ for its Cauchy representation~$\overline  \xi_C$.
\end{thm}

For $i\in\IN$ and $A\in\AA$ let $\mu_i(A):=\mu(F_i\cap A)$. Then $\mu_i$ is a computable measure on
$(\Omega,\AA,\RR,\alpha)$ such that $\mu_i(\Omega)=\mu(F_i)$ is (finite and) $\rho$-computable (see Lemma~\ref{l10} and Theorem~\ref{t4}). For every $i$, $d_i$ defined by $d_i(A,B):=\mu_i(A\syd B)=\mu(F_i\cap(A\syd B))=d(F_i\cap A,F_i\cap B)$ is a computable pseudometric on $\AA$ (not only on $\AA^f$). Notice that $\mu_i$ is the restricton of the measure $\mu$ to $F_i$ and $d_i(A,B)$ is the finite distance of $A$ and $B$ restricted to $F_i$.

Define $e:[0;\infty)\to [0;1)$ by $e(x):=x/(1+x)$. Then $e^{-1}(y)=y/(1-y)$ and $e$ and $e^{-1}$ are  $(\rho,\rho)$-computable increasing functions such that
\begin{eqnarray}
\label{f20}
e(x)\leq x \  \mbox{ and } \  e^{-1}(y)\leq 2\cdot y \mbox{ for } \ y\leq 1/2\,.
\end{eqnarray}
It is known that for a pseudometric $d$,  $d':=e\circ d=d/(1+d)$ is a pseudometric bounded by $1$ with the same induced topology. Furthermore, for a sequence $(d_i)_{i\in\IN}$ of pseudometrics bounded by $1$, $d(x,y):=\sum_{i\in\IN}d_i(x,y)\cdot 2^{-i}$ is a pseudometric \cite{Sch97b}. The statements hold accordingly for metrics.\\

\proof
By the above remarks $\overline d$ is a pseudometric on $\AA$, and since \\
\hspace*{3ex}$\overline d([A],[B])=0$ iff $(\forall i)\,\mu(F_i\cap(A\syd B))=0$ iff $\mu(A\syd B)=0$ iff $[A]=[B]$,\\ $\overline d$ is a metric on $[\AA]$.
Since union, intersection and difference on $\RR$ are $(\alpha,\alpha,\alpha)$-computable, the restriction of $\overline d$ to $\RR$ is $(\alpha,\alpha,\rho)$-computable. Below, we show that $\RR$ is dense in $(\AA,\overline d)$.
\medskip

{\boldmath $\overline \zeta\leq \overline\xi_C$:}
Suppose $\overline\zeta(p)=[A]$. Then $p$ is (encodes) a list of all  $(u,i,v,w)$ such that $\nu_\IQ(u)<\mu(\alpha(v)\cap F_i\cap A)<\nu_\IQ(w)$. From $(p,i)$ we can compute a list $p'$ of all $(u,v,w)$ such that $\nu_\IQ(u)<\mu(\alpha(v)\cap F_i\cap A)<\nu_\IQ(w)$, hence $\zeta(p')=F_i\cap A$.
Since $\zeta\leq\zeta_+$, by Lemma~\ref{l10} we can compute a $\overline \rho_<$-name of $\mu( F_i\cap A)$, hence a $\rho_<$-name of $\mu( F_i\cap A)$ since $\mu( F_i\cap A)$ is finite. Since $\zeta\leq\zeta_-$, by Definition~\ref{d7} we can compute a $\xi_-$-name $q$ of $[F_i\cap A]$.
Therefore, by Theorem~\ref{t4} from $(p,i)$ we can compute a $\xi_C$-name $r$ of $[F_i\cap A]$. Then $r$ is (encodes) a sequence $v_0,v_1,\ldots$ such that
$d(\alpha(v_k),F_i\cap A)\leq 2^{-k}$.

Let $k\in\IN$. Since the metric $d$ is $(\xi,\alpha,\rho)$-computable and $\RR$ is dense in $(\AA^f,d)$, for every $i$ we can find some $u_i$ such that  for $S_i:=\alpha(u_i)$, \\
$\mu((F_i\cap A)\syd S_i)=d(F_i\cap A,S_i)<2^{-k-1}/(k+2)$.
Let $R:=\bigcup_{i\leq k+1}(F_i\cap S_i)$.
Since $D\syd (F\cap S)\In D\syd S$,
\begin{eqnarray*}\mu(F_i\cap(A\syd R))&=&\mu((F_i\cap A)\syd (F_i\cap R))\ =\
\mu((F_i\cap A)\syd (F_i\cap S_i))\\
&\leq& \mu( (F_i\cap A)\syd S_i)
 \ < \  2^{-k-1}/(k+2)
\end{eqnarray*}
and hence
\begin{eqnarray*}
\overline d (A,R)& \leq &
\sum _{i\leq k+1}\frac{\mu(F_i\cap(A\syd R))}{1+\mu(F_i\cap(A\syd R))}\cdot 2^{-i} + 2^{-k-1}\\
& < & (k+2)\cdot \frac{2^{-k-1}}{k+2} +  2^{-k-1}\leq 2^{-k}\,.
\end{eqnarray*}
This implies that  $\RR$ is dense in $(\AA,\overline d)$.
Let $\overline \zeta(p)=[A]$. By Definition~\ref{d9}, for any $i$, an $\alpha$-name of $F_i$ can be computed. So an $\alpha$-name of $R$ can be computed from $p$. Hence a sequence $(v_0,v_1,\ldots)$ can be computed such that
$\overline d(A,\alpha(v_k))\leq 2^{-k-1}$, which by definition constitutes  a
$\overline \xi_C$-name of $[A]$. Therefore $\overline \zeta\leq \overline\xi_C$.

By density of $\RR$, $(\AA,\overline d,\RR,\alpha)$ is a computable pseudometric space and  $([\AA],\overline d,[\RR],\beta)$ is a computable metric space.\\

{\boldmath $\overline \xi_C\leq \overline \zeta$:} We apply the following characterization which is similar to Lemma~\ref{l8}(\ref{l8c}):
\begin{eqnarray}\label{f19}
\hspace*{-3ex}\gamma &\leq & \overline\zeta \iff ([A],R,H)\mapsto \mu(A\cap R\cap H) \ \mbox{ is }\ (\gamma,\alpha,F,\rho)\mbox{-computable}
\end{eqnarray}
Suppose $\overline \xi_C(p)=[A]$, $R=\alpha(v)$ and $H=F_i$. Let $k\in\IN$.
Since $\RR$ is dense in $(\AA,\overline d)$, there is some $S\in\RR$  such that
$\overline d(A,S)\leq 2^{-k-i-1}$. Some $u$ such  that $\overline d(A,S)\leq 2^{-k-i-1}$
for $S=\alpha(u)$ can be computed from $p,i$ and $k$.
\medskip

Since $\mu(F_i\cap R\cap(A\syd S))\leq \mu(F_i\cap(A\syd S))$ and the function $e:x\mapsto x/(1+x)$ is increasing,
\[\sum_j2^{-j}\cdot\frac{\mu(F_j\cap R\cap(A\syd S))}{1+\mu(F_j\cap R\cap(A\syd S))}
 \leq  \sum_j2^{-j}\cdot\frac{\mu(F_j\cap(A\syd S))}{1+\mu(F_j\cap(A\syd S))}
 \leq \overline d(A,S)\,, \]
hence by (\ref{f20}), $\mu(F_i\cap R\cap(A\syd S))\leq 2^{-k}$. It follows that for every $i$,
$\mu((F_i\cap R\cap A)\syd (F_i\cap R\cap S))\leq 2^{-k}$, hence
$|\mu(F_i\cap R\cap A)-\mu (F_i\cap R\cap S)|\leq 2^{-k}$  by (\ref{f16}). Since intersection on $\RR$ is computable, from $A=\overline\xi_C(p)$, $i$, $k$ and $R=\alpha(v)$ we can compute some $a:=\mu (F_i\cap R\cap S)$ such that
$|\mu(F_i\cap R\cap A)-a|\leq 2^{-k}$. Therefore,
$([A],R,H)\mapsto \mu(A\cap R\cap H)$  is $(\overline\xi_C,\alpha,F,\rho)$-computable. By (\ref{f19}), $\overline\xi_C\leq\overline\zeta$.
\qq

\begin{cor}\label{c1} Define $([\AA],\overline d)$ and the Cauchy representation $\overline\xi_C$ as in Theorem~\ref{t5} by a majorising partition $F$ for the notation $\alpha$ of the ring $\RR$. Then
$\zeta\equiv \overline \xi_C$.
\end{cor}

\proof This follows from Theorems~\ref{t9} and~\ref{t5}.
\qq

In the proof of Lemma~\ref{l11} we have constructed a majorizing partition $F$ for $\alpha$.
Although the metric $\overline d$ on $[\AA]$ and the representation $\overline\xi_C$
introduced in Definition~\ref{t9} depend on $F$, the equivalence class of $\overline\xi_C$
is the same for all such partitions.

\section{Summary and final remarks}\label{secg}

Up to equivalence we have the four new representations $\zeta_+,\ \zeta_-,\ \zeta$ and $\xi_C$. The representations $\overline\zeta_+,\ \overline\zeta_-$ and $\overline \zeta$ are equivalent to the first three ones if they are defined by means of a majorising partition which always exists. For the Cauchy representation $\xi_C$ of the sets of finite measure, $\xi_C\equiv \zeta$, if $\mu(\Omega)$ is (finite and) $\rho$-computable. If the Cauchy representation
$\overline \xi_C$ is defined by means of a majorising partition, then $\overline\xi_C\equiv \zeta$.

In \cite{WD05,WD06} Wu and Ding have introduced several other representations of
the measurable sets.
First, we consider \cite{WD06}. The representation $\delta_{\mathbb T_1}$ \cite[Theorem~4.1]{WD06} can be expressed informally as follows:
$\delta_{\mathbb T_1}(p)=[A]$ iff $p$ consists of a list of all
pairs $(E,r)$ such that $\mu(E\setminus A)< r$ and a list of all pairs $(E,r)$ such
that $\mu(A\setminus E)< r$ (where $E\in\RR$ and $r\in\IQ$).
Since $\mu(E)=\mu(E\setminus A) + \mu(E\cap A)$ and $\mu(E)$
can be computed, the first list can be replaced by a list of all pairs $(E,r)$ such that $r <\mu(E\cap A)$.

Define $\delta_1\langle p,q\rangle=[A]$ iff $\zeta_+(p)=[A]$ and $\overline  \rho_>(q)=\mu(A)$.
Then $\delta_1\equiv \delta_{\mathbb T_1}$ (without proof).
Therefore, the restriction of $\delta_{\mathbb T_1}$ to the sets of infinite measure is equivalent to $\zeta_+$ and its restriction to the sets of finite measure is equivalent to
$\xi_+$, hence also equivalent to $\xi_-$, $\xi$ and $\xi_C$ by Theorem~\ref{t4}.

Accordingly, the representation $\delta_{\mathbb T_2}$ from Section~4.2 is equivalent to
the following representation $\delta_2$ defined by
$\delta_2\langle p,q,r\rangle=[A]$ iff $\zeta(p)=[A]$, $\overline\rho_>(q)=\mu(A)$
and $\overline\rho_>(r)=\mu(A^c)$.

The third representation $\delta_{\mathbb T_3}$ from \cite[Section~4.3]{WD06} uses a computable sequence $(C_i)_{i\in\IN}$ where $C_n=\bigcup_{i<n}D_i$ for some partition $(D_i)_{i\in\IN}$ for $\alpha$ such that $\mu(D_i)>0$. The condition  $\mu(D_i)>0$ excludes  some spaces from consideration. It is irrelevant for the representaion $\delta_{\mathbb T_3}$ but important for the representaton $\delta_{\mathbb D_1}$ below. The representation $\delta_{\mathbb T_3}$ can be defined informally as follows:
 $\delta_{\mathbb T_3}(p)=[A]$  iff $p$ is a list of all $(E,i)$  ($E\in\RR$, $i \in\IN$) such that $\mu((A\syd E)\cap C_i)<2^{-i}$.

From $p$ we can compute a list of all $(E,k,r)$ ($r$ rational) such that $\mu((A\syd E)\cap D_k)<$r.
Using arguments similar to those in the proof of Theorem~\ref{t5} we can prove
$\delta_{\mathbb T_3}\equiv\overline\zeta$.
The additional condition $\mu(D_i)>0$ in \cite[Theorem~3.3]{WD06}
is not used in this proof. If the partition $D$ is majorising then $\delta_{\mathbb T_3}\equiv\zeta$ (without proof).

In \cite[Definiton~5.1]{WD05} a metric on $[\AA]$ is defined by
\[d_1([A],[B]):=\sum_{i\in\IN}\frac{\mu(D_i\cap(A\syd B))}{\mu(D_i)}\cdot 2^{-i}\,.\]
This definition is only meaningful if $\mu(D_i)>0$ for all $i$. Therefore, for the metric $\overline d$ in (\ref{t5}) we use  the denominators
$ 1+\mu(D_i\cap(A\syd B))$ instead of $\mu(D_i)$. The
Cauchy representation for the computable metric space
${\mathbb D}_1:=([\AA],d_1,[\RR],[\alpha])$ is called $\delta_{\mathbb D_1}$.
By a proof similar to that of Theorem~\ref{t5} it can be shown that
$\delta_{\mathbb D_1}\equiv \overline\zeta$. By Lemma~\ref{l11} there is a majorising partition $D$. In this case, $\delta_{\mathbb D_1}\equiv \zeta$   by Theorem~\ref{t9}. Also for another metric a Cauchy representation $\delta_{\mathbb D_2}$ is introduced.

Only for the representations  $\delta_{\mathbb T_3}$ and $\delta_{\mathbb D_1}$, which are equivalent (without proof) union and intersection on the measurable sets are computable.
It can be shown that union and intersection are  computable also  for   $\zeta_+$ and $\zeta_-$ and  that countable union is computable for $\zeta_+$ but not for $\zeta$.

A function $f:\Omega\to X$ to a topological space $X$ is measurable, if $f^{-1}(U)$ is measurable for every open set $U$. Since intersection and countable union are computable on the open subsets of a computable topological space \cite{WG09} these operations should also be computable on the measurable sets (since, for example, $f^{-1}(\bigcup U_i)=\bigcup_if^{-1}(U_i)$). From all the representations of measurable sets mentioned in this article only for the representation $\zeta_+$ intersection and countable union are computable.
Therefore, we claim that $\zeta_+$ is the most useful one for studying computability of measurable functions.

In \cite[Sections~4.1 and 4.2]{WD06}  proper supersets
of $\sigma:=\{ \uparrow(E,r)\mid R\in\RR, r\in \IQ_+\}$ where
$\uparrow(E,r):=\{A\in\AA\mid \mu(R\setminus A)<r\}$ have been used as subbases
of topologies for defining the representations $\delta_{\mathbb T_1}$ and $\delta_{\mathbb T_2}$ of the measurable sets. The set $\sigma$ itself would yield a representation  which is equivalent to $\zeta_+$. The authors have not taken this case into consideration.

A representation $\delta\pf \om\to X$ of a topological $T_0$-space $(X,\tau)$ is admissible, iff it is continuous and $\gamma\leq\delta$ for every other continuous representation $\gamma$ of $X$ \cite{Wei00,Sch02,Sch02b,Sch03,BHW08}. For admissible representations, a function on the represented sets is continuous, iff it can be realized by a continuous function on the names.

The Cauchy representation of  a computable metric space is admissible \cite{Wei00}. Therefore, the representations $\xi_C$ (Lemma~\ref{l2}), $\overline \xi_C$ (Theorem~\ref{t5}) and $\delta_{\mathbb D_1}$ \cite{WD05} are admissible.

Let $\lambda:\s\to \mathcal \sigma$ be a notation of a set of subsets of $X$ such that $\sigma$ is a subbase of a $T_0$-topology $(X,\tau)$.
Define a representation $\delta\pf \om\to X$  as follows:
$\delta(p)=x$ iff $p$ is a list of all $w$ such that $x\in\lambda(w)$.
Then $\delta$ is an admissible representation of the space $(X,\tau)$ where $\tau$ is the final topology of $\delta$ \cite{WG09}.
All the other representations of measurable sets defined in this article can be written in this way and hence are admissible. In each case a subbase of the final topology can be directly extracted from the definition.
For example the final topology of $\zeta$ is generated by the subbase consisting of all sets
$B(a,R,b):=\{[A]\in[\AA]\mid a<\mu(R\cap A)<b\}$ such that $a,b\in\IQ$ and $R\in\RR$.

\section{Thanks}
The authors thank the unknown referees for their careful work.

\bibliographystyle{plain}

\begin{thebibliography}{10}

\bibitem{AFR11}
Nathaniel~L. Ackerman, Cameron Freer~E., and Daniel~M. Roy.
\newblock On the computability of conditional probability.
\newblock arXiv:1005.3014, 2011.

\bibitem{Bau74}
Heinz Bauer.
\newblock {\em Wahrscheinlichkeitstheorie und {G}rundz\"uge der
  {M}a\ss{}theorie}.
\newblock Walter de Gruyter, Berlin, 2. edition, 1974.

\bibitem{BDHMS12}
Laurent Bienvenu, Adam Day, Mathieu Hoyrup, Ilya Mezhirov, and Alexander Shen.
\newblock A constructive version of {B}irkhoff's ergodic theorem for
  {M}artin-{L}\"of random points.
\newblock {\em Information and Computation}, 210:21--30, 2012.

\bibitem{BGHRS11}
Laurent Bienvenu, Peter G{\'a}cs, Mathieu Hoyrup, Rojas Crist{\'o}bal, and
  Alexander Shen.
\newblock Algorithmic tests and randomness with respect to a class of measures.
\newblock {\em Proceedings of theSteklov Institute of Mathematics (Trudy
  Miran)}, 270(1):34--89, 2011.

\bibitem{Bos08b}
Volker Bosserhoff.
\newblock Notions of probabilistic computability on represented spaces.
\newblock {\em Journal of Universal Computer Science}, 14(6):956--995, 2008.

\bibitem{BG11a}
Vasco Brattka and Guido Gherardi.
\newblock Effective choice and boundedness principles in computable analysis.
\newblock {\em The Bulletin of Symbolic Logic}, 17(1):73--117, 2011.

\bibitem{BG11}
Vasco Brattka and Guido Gherardi.
\newblock Weihrauch degrees, omniscience principles and weak computability.
\newblock {\em The Journal of Symbolic Logic}, 76(1):143--176, 2011.

\bibitem{BHW08}
Vasco Brattka, Peter Hertling, and Klaus Weihrauch.
\newblock A tutorial on computable analysis.
\newblock In S.~Barry Cooper, Benedikt L\"owe, and Andrea Sorbi, editors, {\em
  New Computational Paradigms: Changing Conceptions of What is Computable},
  pages 425--491. Springer, New York, 2008.

\bibitem{BP03}
Vasco Brattka and Gero Presser.
\newblock Computability on subsets of metric spaces.
\newblock {\em Theoretical Computer Science}, 305:43--76, 2003.

\bibitem{Eda95}
Abbas Edalat.
\newblock Domain theory and integration.
\newblock {\em Theoretical Computer Science}, 151:163--193, 1995.

\bibitem{Eda09}
Abbas Edalat.
\newblock A computable approach to measure and integration theory.
\newblock {\em Information and Computation}, 207(5):642--659, 2009.

\bibitem{Fol99}
Gerald~B. Folland.
\newblock {\em Real Analysis: Modern Techniques and Their Applications}.
\newblock Wiley, 2 edition, 1999.

\bibitem{FR12}
Cameron~E. Freer and Daniel~M. Roy.
\newblock Computable {F}inetti measures.
\newblock {\em Annals of Pure and Applied Logic}, 163:530--546, 2012.
\newblock arXiv:0912.1072.

\bibitem{Gac05}
Peter G{\'{a}}cs.
\newblock Uniform test of algorithmic randomness over a general space.
\newblock {\em Theoretical Computer Science}, 341:91--137, 2005.

\bibitem{GHR09b}
Peter Ga\'cs, Mathieu Hoyrup, and Crist\'obal Rojas.
\newblock Randomness on computable probability spaces-a dynamical point of
  view.
\newblock {\em Theory of Computing Systems, special issue STACS 09}, 2010.

\bibitem{GHR10}
Stefano Galatolo, Mathieu Hoyrup, and Crist\'obal Rojas.
\newblock Effective symbolic dynamics, random points, statistical behavior,
  complexity and entropy.
\newblock {\em Information and Computation}, 208(1):23--41, 2010.

\bibitem{GHR11}
Stefano Galatolo, Mathieu Hoyrup, and Crist{\'o}bal Rojas.
\newblock Dynamics and abstract computability: computing invariant measures.
\newblock {\em Discrete and Continuous Dynamical Systems - Series A},
  29:193--212, 2011.

\bibitem{Hoy13}
Mathieu Hoyrup.
\newblock Computability of the ergodic decomposition.
\newblock {\em Annals of Pure and Applied Logic}, 164(5):542--549, 2013.

\bibitem{HR09}
Mathieu Hoyrup and Christ{\'o}bal Rojas.
\newblock Computability of probability measures and {M}artin-{L}{\"o}f
  randomness over metric spaces.
\newblock {\em Information and Computation}, 207:830--847, 2009.

\bibitem{HRW11c}
Mathieu Hoyrup, Crist\'obal Rojas, and Klaus Weihrauch.
\newblock Computability of the {R}adon-{N}ikodym derivative.
\newblock {\em Computability}, 1(1):3--13, 2012.

\bibitem{MTY13}
Takakazu Mori, Yoshiki Tsujii, and Mariko Yasugi.
\newblock computability of probability distributions and characteristic
  functions.
\newblock {\em Logical Methods in Computer Science}, 9:1--11, 2013.

\bibitem{Mue99}
Norbert~Th. M{\"u}ller.
\newblock Computability on random variables.
\newblock {\em Theoretical Computer Science}, 219:287--299, 1999.

\bibitem{Sch97b}
Eric Schechter.
\newblock {\em Handbook of Analysis and Its Foundations}.
\newblock Academic Press, San Diego, 1997.

\bibitem{Sch02b}
Matthias Schr{\"o}der.
\newblock Effectivity in spaces with admissible multirepresentations.
\newblock {\em Mathematical Logic Quarterly}, 48(Suppl. 1):78--90, 2002.

\bibitem{Sch02}
Matthias Schr{\"o}der.
\newblock Extended admissibility.
\newblock {\em Theoretical Computer Science}, 284(2):519--538, 2002.

\bibitem{Sch03}
Matthias Schr\"oder.
\newblock Admissible representations for continuous computations.
\newblock Informatik Berichte 299, FernUniversit\"at Hagen, Hagen, April 2003.
\newblock Dissertation.

\bibitem{Sch07b}
Matthias Schr{\"o}der.
\newblock Admissible representations for probability measures.
\newblock {\em Mathematical Logic Quarterly}, 53(4--5):431--445, 2007.

\bibitem{Wei92a}
Klaus Weihrauch.
\newblock The degrees of discontinuity of some translators between
  representations of the real numbers.
\newblock Technical Report TR-92-050, International Computer Science Institute,
  Berkeley, July 1992.

\bibitem{Wei99a}
Klaus Weihrauch.
\newblock Computability on the probability measures on the {B}orel sets of the
  unit interval.
\newblock {\em Theoretical Computer Science}, 219:421--437, 1999.

\bibitem{Wei00}
Klaus Weihrauch.
\newblock {\em Computable Analysis}.
\newblock Springer, Berlin, 2000.

\bibitem{WG09}
Klaus Weihrauch and Tanja Grubba.
\newblock Elementary computable topology.
\newblock {\em Journal of Universal Computer Science}, 15(6):1381--1422, 2009.

\bibitem{WWD09}
Klaus Weihrauch, Yongcheng Wu, and Decheng Ding.
\newblock Absolutely non-computable predicates and functions in analysis.
\newblock {\em Mathematical Structures in Computer Science}, 19:59--71, 2009.

\bibitem{Wu12}
Yongcheng Wu.
\newblock Computability on random events and variables in a computable
  probability space.
\newblock {\em Theoretical Computer Science}, 460:54--69, 2012.

\bibitem{WD05}
Yongcheng Wu and Decheng Ding.
\newblock Computability of measurable sets via effective metrics.
\newblock {\em Mathematical Logic Quarterly}, 51(6):543--559, 2005.

\bibitem{WD06}
Yongcheng Wu and Decheng Ding.
\newblock Computability of measurable sets via effective topologies.
\newblock {\em Archive for Mathematical Logic}, 45(3):365--379, 2006.

\bibitem{WW06}
Yongcheng Wu and Klaus Weihrauch.
\newblock A computable version of the {D}aniell-{S}tone theorem on integration
  and linear functionals.
\newblock {\em Theoretical Computer Science}, 359(1--3):28--42, 2006.

\end{thebibliography}

\section{Appendix: Some useful rules for the symmetric difference}\label{sech}

\begin{eqnarray}
\label{f5}A\syd B &=& B\syd A,\\
\label{f6}(A\syd B)\syd C&=& A\syd (B\syd C),\\
\label{f7}A\syd B&\In &A\syd C \cup C\syd B, \\
\label{f8}A\cup B&=& A\cap B \uplus A\syd B,\\
\label{f9}A&\In& B\cup (A\syd B),\ \\
\label{f10}(A\syd B)\cap C&=&(A\cap C)\syd(B\cap C)=(C\setminus A)\syd(C\setminus B), \\
\label{f11}(\bigcup_{i\in I} A_i)\  \syd \ (\bigcup_{i\in I} B_i) &\In& \bigcup_{i\in I} (A_i\syd B_i)\,.
\end{eqnarray}

Let $\mu$ be a measure on a ring $\RR$. From (\ref{f8}),
$$\mu(A)\leq \mu(A\cup B)=\mu(A\cap B) +\mu(A\syd B)\leq \mu(B)+\mu(A\syd B)$$
and accordingly with $A$ and $B$ interchanged. Therefore,
\begin{eqnarray}
\label{f3}&
 \mu(A) \leq  \mu(B) +\mu(A\syd B)\,,\\
\label{f1}&
  \mu(A)=\mu(A\cup B)=\mu(A\cap B)=\mu(B) \ \ \mbox{if}\  \mu(A\syd B)=0\,,\\
 \label{f16}&
| \mu(A)- \mu(B)| \leq  \mu(A\syd B)\ \ \ \mbox{if $A$ and $B$ have finite measure}\,.
\end{eqnarray}

\section{Appendix: Proof of Theorem~\ref{t2}}\label{seci}

\noindent By (\ref{f7}) the mapping $d:(A,B)\mapsto
\mu(A\syd B)$ is a pseudometric on the set ${\AA}^f$.
\medskip

Next we prove (\ref{t2b}).\\
Obviously, $B_{mk}\In B_{m,k+1}$.
Since $(X\cup Y\cup Z)\setminus (X\cup Y)\In Y\syd Z$,
\begin{eqnarray*} d(B_{mk},B_{m,k+1})
 &=& \mu((A_m\cup\ldots \cup A_k\cup A_{k+1})\setminus(A_m\cup\ldots \cup A_k))\\
 &\leq& \mu(A_k\syd A_{k+1})\leq 2^{-k}
\end{eqnarray*}
and  $d(B_{mk}, B_{mk'})
\leq 2^{-k}+\ldots +2^{-(k'-1)}<2\cdot 2^{-k}$ for $k<k'$ by induction. Therefore,
$d(B_{mk},B_m)= \mu(B_m\setminus B_{mk})=\mu((\bigcup_{k< k'} B_{mk'})\setminus B_{mk}) =\mu(\bigcup_{k< k'} (B_{mk'}\setminus B_{mk}))=
\sup_{k<k'}\mu(B_{mk'}\setminus B_{mk})\leq  2\cdot 2^{-k}$. This proves (\ref{f21}).
\smallskip

Obviously $B_m \supseteq B_{m+1}$ . Since $B_{mm}\in \AA^f$ and $d(B_{mm},B_m)$ is finite, $B_m\in\AA^f$.
Since $(X\cup Y\cup Z)\setminus (X\cup Y)\In Y\syd Z$,
\begin{eqnarray*}
d(B_m,B_{m+1}) & = & \mu ( (A_m \cup A_{m+1} \cup\ldots)\setminus
(A_{m+1} \cup A_{m+2} \cup \ldots))\\
&\leq & \mu(A_m\syd A_{m+1})\leq 2^{-m}
\end{eqnarray*}
and  $d(B_{m}, B_{m'})
\leq 2^{-m}+\ldots +2^{-(m'-1)}<2\cdot 2^{-m}$ for $m<m'$ by induction.
Therefore,
$d(B_m,B)= \mu(B_m\setminus B)=\mu(B_m\setminus \bigcap_{m<m'}B_{m'})
=\mu(\bigcup_{m<m'}(B_m\setminus B_{m'}))
=\sup_{m<m'}\mu(B_m\setminus B_{m'})\leq 2\cdot 2^{-m}$.
This proves (\ref{f22}).
\smallskip

$B\in \AA^f$ since $B\In B_0$ and $B_0\In \AA^f$. By (\ref{f21}, \ref{f22}),
$d(A_m,B)=d(B_{mm},B)\leq d(B_{mm}, B_m)+ d(B_m, B)\leq 4\cdot 2^{-m}$.
This proves (\ref{f23}).

\medskip
Next we prove (\ref{t2d}).
Obviously, $D_{mk}\supseteq D_{m,k+1}$ .
Since $(X\cap Y)\setminus (X\cap Y\cap Z)\In Y\syd Z$,
\begin{eqnarray*}
d(D_{mk},D_{m,k+1}) & = & \mu((A_m\cap\ldots \cap A_{k})\setminus (A_m\cap\ldots \cap A_{k}\cap A_{k+1}))\\
 &\leq & \mu(A_k\syd A_{k+1})\leq 2^{-k}\,,
 \end{eqnarray*}
and  $d(D_{mk}, D_{mk'})
\leq 2^{-k}+\ldots +2^{-(k'-1)}<2\cdot 2^{-k}$ for $k<k'$ by induction.
Therefore,
$d(D_{mk},D_m)= \mu(D_{mk}\setminus D_m)
=\mu(D_{mk}\setminus(\bigcap_{k< k'} D_{mk'}))
=\mu( \bigcup_{k<k'}(D_{mk}\setminus D_{mk'}))
 = \sup_{k<k'}\mu(D_{mk}\setminus D_{mk'})\leq  2\cdot 2^{-k}$.
This proves (\ref{f25}).

\smallskip
Obviously, $D_m\In D_{m+1}$. $D_m\in\AA^f$ since $D_{mm}\in \AA^f$ and $D_m\In D_{mm}$.\\
Since $(X\cap Y)\setminus (Z\cap X\cap Y)\in Y\syd Z$,
\begin{eqnarray*}
d(D_m,D_{m+1}) &=& \mu(
(A_{m+1}\cap A_{m+2}\cap\ldots)\setminus (A_m\cap A_{m+1}\cap\ldots)) \\
&\leq &\mu( A_m\syd A_{m+1})\leq 2^{-m}
\end{eqnarray*}
and  $d(D_{m}, D_{m'})\leq 2^{-m}+\ldots +2^{-(m'-1)}<2\cdot 2^{-m}$
for $m<m'$ by induction. Therefore,
$d(D_m,D)= \mu(D\setminus D_m)=\mu((\bigcup_{m<m'} D_{m'})\setminus D_m)
=\mu(\bigcup_{m<m'}(D_{m'}\setminus D_m))
=\sup_{m<m'}\mu(D_{m'}\setminus D_m)\leq 2\cdot 2^{-m}$.
This proves (\ref{f26}).

\smallskip
$D\in \AA^f$ since $D_0\in \AA^f$ and $d(D,D_0)$ is finite. By (\ref{f25}, \ref{f26}),
$d(A_m,D)=d(D_{mm},D)\leq d(D_{mm}, D_m)+ d(D_m, D)\leq 4\cdot 2^{-m}$.
This proves (\ref{f27}). Altogether we have proved  (\ref{t2d}).
\medskip

From (\ref{t2b}) or (\ref{t2d}) it follows that $(\AA^f,d)$ is a complete pseudometric space.
\medskip

We prove (\ref{t2c}), i.e. density of $\RR^f$. For $C\In \Omega$ let ${\mathcal U}(C)$ be the set of all sequences $(R_i)_{i\in\IN}$ of ring elements such that $C\In\bigcup _{i\in\IN}R_i$. In the Carath\'eodory proof of the extension theorem \cite{Bau74} the measure $\mu$ is defined on $\AA$ by its values on the ring as follows:
\[ \mu(C):=\inf\{\sum_{i\in\IN}\mu(R_i)\mid (R_i)_{i\in\IN}\in {\mathcal U}(C)\}\,.\]

Let $C\in\AA^f$ and let $\varepsilon >0$. There is some sequence $(R_i)_{i\in\IN}\in {\mathcal U}(C)$ such that $C\In\bigcup _{i\in\IN}R_i$ and $0\leq \sum_{i\in\IN}\mu(R_i)-\mu(C)<\varepsilon/2$. Then $(\forall i)\,R_i\in\RR^f$.
Let $S_0:= R_0$ and $S_i:= R_i\setminus (R_0\cup\ldots\cup R_{i-1})$
for all $i>0$. Then the $S_i$ are pairwise disjoint sets of finite measure and $C\In\bigcup _{i\in\IN}R_i =\bigcup _{i\in\IN}S_i$.
 Since  $S_i\In R_i$ for all $i$,
\[\mu( C\syd \bigcup _{i\in\IN}S_i)=\mu( C\syd \bigcup _{i\in\IN}R_i)
= \mu(\bigcup_{i\in\IN}R_i)-\mu(C)\leq \sum _{i\in\IN}\mu(R_i)-\mu(C)
\leq\varepsilon/2\,.
\]
Furthermore there is some $m$ such that
$0\leq \sum_{i\in\IN}\mu(S_i)-
\sum_{i\leq m}\mu(S_i)<\varepsilon/2$. Since the $S_i$ are disjoint,  $0\leq \mu( \bigcup_{i\in\IN}S_i)-
\mu(\bigcup_{i\leq m}S_i)<\varepsilon/2$, hence

\[\mu( \bigcup_{i\in\IN}S_i  \syd \bigcup_{i\leq m}S_i)=
\mu( \bigcup_{i\in\IN}S_i)-\mu(\bigcup_{i\leq m}S_i)<\varepsilon/2\,.\]
By (\ref{f7}), $\mu(C\syd \bigcup_{i\leq m}S_i)\leq
\mu( C\syd \bigcup _{i\in\IN}S_i)+ \mu(\bigcup_{i\in\IN}S_i\syd \bigcup_{i\leq m}S_i)\leq\varepsilon$. Since $\bigcup_{i\leq m}S_i\in \RR^f$,
$\RR^f$ is dense in $\AA^f$.
\qq

\end{document}